\documentclass[twocolumn]{aastex631}

\usepackage{textcomp}
\usepackage{subfigure}
\usepackage{appendix}

\usepackage{comment}
\usepackage{rotating}
\usepackage{amsmath}
\usepackage{amssymb}
\usepackage{amsfonts}

\newcommand{\x}{\mbox{$\times$}}

\shorttitle{AASTeX v6.3.1}
\shortauthors{Morii et al.}

\begin{document}

\title{Global and Local Infall in the ASHES Sample (GLASHES). I. Pilot Study in G337.541} 

\correspondingauthor{Kaho Morii}
\email{kaho.morii@grad.nao.ac.jp}

\author[0000-0002-6752-6061]{Kaho Morii}
\affil{Department of Astronomy, Graduate School of Science, The University of Tokyo, 7-3-1 Hongo, Bunkyo-ku, Tokyo 113-0033, Japan}
\affil{National Astronomical Observatory of Japan, National Institutes of Natural Sciences, 2-21-1 Osawa, Mitaka, Tokyo 181-8588, Japan} 

\author[0000-0002-7125-7685]{Patricio Sanhueza}
\affil{Department of Earth and Planetary Sciences, Institute of Science Tokyo, Meguro, Tokyo, 152-8551, Japan}
\affil{National Astronomical Observatory of Japan, National Institutes of Natural Sciences, 2-21-1 Osawa, Mitaka, Tokyo 181-8588, Japan}
\affil{Department of Astronomical Science, SOKENDAI (The Graduate University for Advanced Studies), 2-21-1 Osawa, Mitaka, Tokyo 181-8588, Japan}

\author[0000-0002-6018-1371]{Timea Csengeri}
\affil{Laboratoire d'Astrophysique de Bordeaux, Univ. Bordeaux, CNRS, B18N, all\'{e}e Geoffroy Saint-Hilaire, 33615, Pessac, France}

\author[0009-0007-6357-6874]{Fumitaka Nakamura}
\affil{National Astronomical Observatory of Japan, National Institutes of Natural Sciences, 2-21-1 Osawa, Mitaka, Tokyo 181-8588, Japan} 
\affil{Department of Astronomy, Graduate School of Science, The University of Tokyo, 7-3-1 Hongo, Bunkyo-ku, Tokyo 113-0033, Japan} 

\author[0000-0002-4093-7178]{Sylvain Bontemps}
\affil{Laboratoire d'Astrophysique de Bordeaux, Univ. Bordeaux, CNRS, B18N, all\'{e}e Geoffroy Saint-Hilaire, 33615, Pessac, France}

\author[0000-0003-1649-7958]{Guido Garay}
\affil{Departamento de Astronomía, Universidad de Chile, Las Condes, Santiago 7550000, Chile}
\affil{Chinese Academy of Sciences South America Center for Astronomy, National Astronomical Observatories, CAS, Beĳing 100101, China}

\author[0000-0003-2384-6589]{Qizhou Zhang}
\affiliation{Center for Astrophysics $|$ Harvard \& Smithsonian, 60 Garden Street, Cambridge, MA 02138, USA}

\begin{abstract}
Recent high-angular-resolution observations indicate the need for core growth to form high-mass stars. 
To understand the gas dynamics at the core scale in the very early evolutionary stages before being severely affected by feedback, we have conducted Atacama Large Millimeter/submillimeter Array (ALMA) observations toward a 70 $\mu$m dark massive clump, G337.541-00.082 as part of the Global and Local infall in the ASHES sample (GLASHES) program. 
Using dense gas tracers such as N$_2$H$^+$ ($J = 1-0$) and HNC ($J = 3-2$), we find signs of infall from the position-velocity diagram and more directly from the blue asymmetry profile in addition to the clump-scale velocity gradient. 
We estimate infall velocities from intermediate and low-mass cores to be 0.28--1.45 km\,s$^{-1}$, and infall rates to be on the order of 10$^{-4}$ to 10$^{-3}$\,$M_\odot$\,yr$^{-1}$, both are higher than those measured in low-mass star-forming regions by more than a factor of five and an order of magnitude, respectively. 
We find a strong correlation of the infall velocity with the nonthermal velocity dispersion, suggesting that infall may contribute significantly to the observed line width. 
Consistent with clump-fed scenarios, we show that the mass infall rate is larger for larger core masses and shorter distances to the clump center. 
Such high infall rates in cores embedded in IRDCs can be considered as strong signs of core growth, allowing high-mass star formation from intermediate-mass cores that would not initially form high-mass stars at their current mass.   
\end{abstract}
\keywords{Infrared dark clouds, Star formation, Star forming regions}

\section{Introduction} \label{sec:intro} 
How to accumulate a large amount of gas in a short time scale is one of the key questions to understand the high-mass star formation process. 
Recent high-resolution observations of regions hosting the early stages of high-mass star formation or infrared dark clouds (IRDCs) reveal the lack of high-mass prestellar cores \citep[e.g.,][]{Zhang11, Sanhueza13,Sanhueza17,Lu15, Ohashi16, Louvet19, Pillai19, Morii21},  including the ALMA Survey of 70 $\mu$m Dark High-mass Clumps in Early Stages \citep[ASHES; ][]{Sanhueza19, Morii23}, with few exceptions in more evolved regions \citep[][]{Kong17, Nony18, Molet19, Barnes23, Mai24}, suggesting the need for cores to grow in order to form a high-mass star.  
In the ASHES survey, the most massive cores observed in IRDCs have masses of $\sim$10\,$M_\odot$, which requires a high mass infall rate of the order of 10$^{-3}$\,$M_\odot$\,yr$^{-1}$ to form high-mass stars, assuming a core-to-star formation efficiency of 30--50\% and an accretion time scale of around a free-fall time scale (a few $\times 10^4$\,yr). 
One traditional method to measure the infall rate is to use the asymmetric double-peak line profile of optically thick lines that emerge due to the line-of-sight velocity and temperature gradient and self-absorption \citep[][and references there in]{De_Vries05}. 
There are many previous studies using this line asymmetry to study infall motions in nearby low-mass cores \citep[e.g., ][]{Lee01, Campbell16, Yu22} and also in high-mass star-forming regions 
\citep[e.g., ][]{Fuller05, Sanhueza10, Schneider10, Csengeri11, Reiter11,  Rygl13, Wyrowski16, Jackson19, Traficante17, Jin-Jin21, Yang21, Xu23_hcn}. 
However, since high-mass star-forming regions are distant, the core-scale estimation, which is likely directly related to the individual star-formation process, is very limited. 
\citet{Contreras18} observed an intermediate mass prestellar core, ALMA1, embedded in the IRDC G331.372-00.116 with the Atacama Large Millimeter/submillimeter Array (ALMA) that includes 12 m, 7 m, and TP arrays. They  
found a blue asymmetry profile and estimated a mass infall rate of 1.96$\times$10$^{-3} M_\odot$\,yr$^{-1}$. 
However, they report only one case, and the infall around other cores remains unexplored. 
To characterize the infall rates in IRDCs and make a comparison with the expected high-mass infall rate to achieve high-mass star formation and with rates in low-mass star-forming regions, a larger sample is needed.  
Furthermore, the velocity gradient traced by dense gas tracers, such as N$_2$H$^+$ and H$^{13}$CO$^+$, has also been used to study the internal kinematic structure of clouds, including gas inflow along filaments \citep[e.g., ][]{Henshaw14, Lu18, Peretto14, Sanhueza21, Redaelli22,Xu23}. 

G337.541–00.082 (hereafter G337), one of the ASHES targets, is a part of an IRDC and is considered to be in the very early evolutionary stage of high-mass star formation without bright infrared point sources at $<$70 $\mu$m, indicating the absence of embedded protostars \citep[see Figure~3 in][]{Sanhueza19}. 
The clump properties have been derived in a series of works by the MALT90 team, and summarized in \citet{Morii23}. 
The kinematic distance of G337 is 4.0 kpc \citep{Whitaker17} using $v_\mathrm{LSR}= -54.6\mathrm{\,km\,s^{-1}}$ \citep{Rathborne16}, and the dust temperature derived from the SED fitting technique is 12.0\,K \citep{Guzman15}. 
G337 is a massive ($M=1200 M_\odot$) and dense ($\Sigma=0.46$ g\,cm$^{-2}$, $n$(H$_2) = 5.49 \times 10^4$\,cm$^{-3}$) clump with a size of $r$$\sim$0.4 pc. 
Previous ALMA observations in the ASHES project revealed nineteen cores from 1.3 mm continuum emission with an angular resolution of $\sim$1.2'' \citep[][also see white contours in Figure~\ref{fig:n2hp_mom}]{Sanhueza19,Morii23,Morii24}. 
Among these cores, two intermediate mass cores ($M=10$ and 4\,$M_\odot$ for ALMA1 and ALMA2, respectively) are found in the center of the clump, both associated with outflow signatures in the east-west direction detected in CO ($J = 2-1$), SiO ($J = 5-4$), and H$_2$CO ($J = 3-2$) \citep{Li21a,Izumi24}. 
Both cores have line detection of deuterated molecules, such as DCO$^+$ ($J = 3-2$), and their estimated virial parameters, without considering the contribution of magnetic fields, are 0.74 and 0.72 \citep{Li22,Li23}. 
Consequently, they are thought to be collapsing and hosting protostars.  
Note that the outflow properties and the absence of 70 $\mu$m point sources imply that the protostars are young, not yet massive. 
Another intermediate mass core, ALMA3 ($M= 5.3\,M_\odot$), is located north of the clump and is also associated with an outflow. 
All remaining cores are low-mass cores ($\lesssim$1\,$M_\odot$) without outflow emission. 

In this paper, we report the pilot study of the clump-scale and core-scale infall motion using N$_2$H$^+$ ($J = 1-0$) and HNC ($J = 3-2$) lines in G337 as part of the Global and Local infall in the ASHES sample (GLASHES) project. 
We describe the observation setup in Section~\ref{sec:obs} and show the observational results of each line in Section~\ref{sec:results}. 
In Section~\ref{sec:analysis}, we analyze the line emission to estimate the infall properties and discuss the estimated physical parameters comparing within the clump and with low-mass star forming regions in Section~\ref{sec:discussion}. 
Section~\ref{sec:conclusion} presents a summary of our work. 

\section{Observations and Data Reduction}
\label{sec:obs}
We use ALMA observations taken during Cycle 6 (2018.1.00299.S, PI:, Y. Contreras). 
The project consists of data from band 3 and 6 taken with the ALMA 12\,m array, the Atacama Compact 7 m array (ACA), and total power (TP). 
The band 3 observations were carried out on 29 December 2018, 31 March, and 1 April 2019 (ALMA array 12\,m), 30 and 31 March 2019 (ACA), and 1, 2, 5, 6, 21, 26, and 29 May 2019 (TP).  
The phase reference center was R.A. (J2000.0) = $16^h 37^m 58 \fs 437$ and Dec (J2000.0) = $-47^\circ 09^\prime 00\farcs 729$. 
The band 6 observations were made on 23 March 2019 (ALMA 12\,m array),  21 and 22 March 2019 (ACA), and 13, 18, and 19 May 2019 (TP). 
The phase reference center was R.A. (J2000.0) = $16^h 37^m 58 \fs 437$ and Dec (J2000.0) = $-47^\circ 09^\prime 00\farcs 330$. 
Both band 3 and 6 observations were single-pointing.  
The ALMA 12\,m array consisted of 46 antennas, with a baseline ranging from 15.1 to 500.2 m (band 3) and 15.1 to 313.7 m (band 6).  
Flux calibration and phase calibration were performed using J1650-5044.
The quasar J1427-4206 or J1617-5848 was used for bandpass calibration. 
The total on-source time was $\sim$24 and 11 minutes, for band 3 and band 6, respectively. 
These observations are sensitive to angular scales smaller than $\sim$26.37$''$ (band 3) and $\sim$9.0$''$ (band 6). 
More extended emission was recovered by including ACA data. 
The 7 m array observations consisted of 11 or 12 antennas, with baselines ranging from 8.9 to 48.9 m. 
Flux and phase calibration were performed using J1650-5044, and bandpass calibration was performed using J1427-4206 (band 3) and J1517-2422 (band 6).
The total on-source time was $\sim$41 and 37 minutes for ACA in band 3 and band 6, respectively.  
These observations are sensitive to angular scales smaller than $\sim$44.7$''$ (band 3) and $\sim$15.3$''$ (band 6). 
We also obtained total power data to cover the zero-baseline information. 
The total power array consisted of three antennas and the total on-source time was 93.5 minutes and 165 minutes for band 3 and band 6, respectively. 

In this paper, we use three molecular lines from these observations, N$_2$H$^+$ ($J = 1-0$) in band 3 and HNC ($J = 3-2$) and HCO$^+$ ($J = 3-2$) in band 6. 
The velocity resolution of N$_2$H$^+$ is $\sim$0.19 km\,s$^{-1}$, and that of HNC and HCO$^+$ is 0.27 km\,s$^{-1}$. 
Data reduction is performed using the CASA software package 5.4.0 for calibration and 5.6.0 for imaging \citep[]{CASA22}. 
We use the automatic cleaning algorithm for imaging data cubes, YCLEAN \citep[][]{Contreas_yclean_18, Contreras18} to CLEAN the 12 m array and the 7 m array combined data cubes for each spectral window with custom-made masks. 
We adopt a Briggs\textquotesingle s robust weighting of 0.5. 
The channel widths used to measure the noise level are $\sim$0.1\,km\,s$^{-1}$ for N$_2$H$^+$ and $\sim$0.14\,km\,s$^{-1}$ for HNC and HCO$^+$, resulting in an average 1$\sigma$ rms noise level measured from line-free channel of 5.0 mJy beam$^{-1}$ (N$_2$H$^+$), 7.0 mJy beam$^{-1}$ (HNC), and 7.5 mJy beam$^{-1}$ (HCO$^+$), respectively. 
Furthermore, we combine the 12m+7m cube with the TP observations using the CASA task ``feather''. 
The synthesized beam sizes of all combined cubes are $2\farcs 27 \times 1\farcs 75$ (-82.3$\arcdeg$) for N$_2$H$^+$ and $1\farcs 05 \times 1\farcs 03$ (-86.3$\arcdeg$) for HNC and HCO$^+$.  
At a source distance of 4.0 kpc, these correspond to $\sim$8000 au and $\sim$4000 au on a linear scale, respectively. 
All images of N$_2$H$^+$, HNC, and HCO$^+$ shown in the paper are ALMA 12\,m, ACA and TP combined before the primary beam correction. 

Furthermore, we also use the 1.3 mm continuum, DCO$^+$ ($J = 3-2$), CO ($J = 2-1$), and SiO ($J = 5-4$) lines produced in previous studies \citep{Sanhueza19, Li20, Li23}. 
These are 12m+7m combined datasets, thus without TP observations. The details of the data reduction are described in the corresponding papers. 
The synthesized beam sizes of the 12m + 7m cube are $1\farcs 14 \times 1\farcs 07$ (79$\arcdeg$) for the continuum, $1\farcs 55 \times 1\farcs 42$ (84$\arcdeg$) for DCO$^+$, $1\farcs 45 \times 1\farcs 34$ (81$\arcdeg$) for CO,  and $1\farcs 55 \times 1\farcs 44$ (86$\arcdeg$) for SiO. 
Table~\ref{tab:spw} shows the summary of the line and the cubes produced.  

\begin{deluxetable*}{lccccc}
\label{tab:spw}
\tabletypesize{\footnotesize}
\tablecaption{Summary of the molecular lines and cube data}
\tablewidth{0pt}
\tablehead{
\colhead{Transition}  & \colhead{Rest Frequency} & \colhead{$E_{\mathrm{u}}/k$} & \colhead{Velocity Resolution} & \colhead{rms noise level} & \colhead{Beam size}  \\ 
\colhead{} & \colhead{GHz} & \colhead{K}  & \colhead{\,km\,s$^{-1}$} & \colhead{mJy beam$^{-1}$} & \colhead{arcsec $\times$ arcsec}} 
\startdata
    N$_2$H$^+$ ($J=1-0$) &  93.173763  & 4.47 & 0.19 & 5.0 & 2.27$\times$1.75\\
    HCO$^+$ ($J=3-2$)    & 267.557633 & 25.68 & 0.27 & 7.0 & 1.05$\times$1.03\\
    HNC ($J=3-2$)        & 271.981111 & 26.10  & 0.27 & 7.5 & 1.05$\times$1.03 \\
    DCO$^+$ ($J=3-2$)    & 216.112580 & 20.74  & 0.17 & 6.5 & 1.55$\times$1.42\\
    CO ($J=2-1$)         & 230.538000 & 16.60  & 1.27 & 2.7 & 1.45$\times$1.34\\
    SiO ($J=5-4$)        & 217.104980 & 31.26  & 0.17 & 6.3 & 1.55$\times$1.44\\
\enddata
\end{deluxetable*}

\section{Results} 
\label{sec:results}

\subsection{N$_2$H$^+$ ($J = 1-0$)}
\begin{figure*}
    \centering
    \includegraphics[width=16cm]{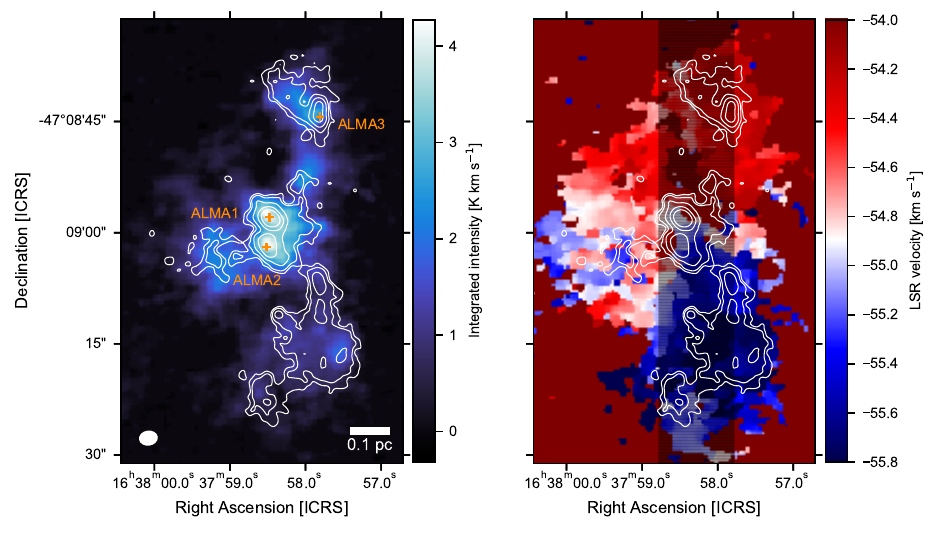}
    \caption{The integrated intensity (mom0) map and the line center map of N$_2$H$^+$ ($J=1-0$, $F_1, F=0,1-1,2$). The white contours show 1.3 mm continuum emission at levels of 3, 5, 10, 20, and 40 sigma, where 1 sigma = 0.07 Jy\,beam$^{-1}$. The synthesized beamsize and the scale bar are plotted in the bottom of the left panel. In the left panel, orange crosses represent the continuum peak positions of ALMA1, ALMA2, and ALMA3. In the right panel, the region used for making position-velocity diagram (Figure~\ref{fig:pv}) is shaded in black. }
    \label{fig:n2hp_mom}
\end{figure*}
To investigate the gas dynamics in the entire clump, we mapped the isolated hyperfine component of N$_2$H$^+$,  $J=1-0$, $F_1, F=0,1-1,2$ transition ($\nu=93.176265$ GHz) because it is isolated, spectral resolved hyperfine component with typically lower optical depth. 
Figure~\ref{fig:n2hp_mom} presents the integrated intensity map (mom0) and the line center map by using the quadratic method \citep{Teague18}. 
The left panel indicates that the spatial distribution of N$_2$H$^+$ emission is almost consistent with the emission of the 1.3\,mm continuum, but the continuum emission is more compact, implying that it traces the densest regions. 
N$_2$H$^+$ traces more extended regions than the continuum-traced parts, for example, it is detected connecting the north gas blob and the central gas blob, which are seen separately in the continuum. 
The line center map indicates a clump-scale velocity gradient from north to south with a velocity variation more than 1 km\,s$^{-1}$, as well as a complex and relatively sharp transition around the central two cores. 
This large-scale gradient is seen to be perpendicular to the outflow directions (east-west) from the central two cores, implying that this gradient is unrelated to the molecular outflows. 

\begin{figure*}
    \centering
    \includegraphics[width=16cm]{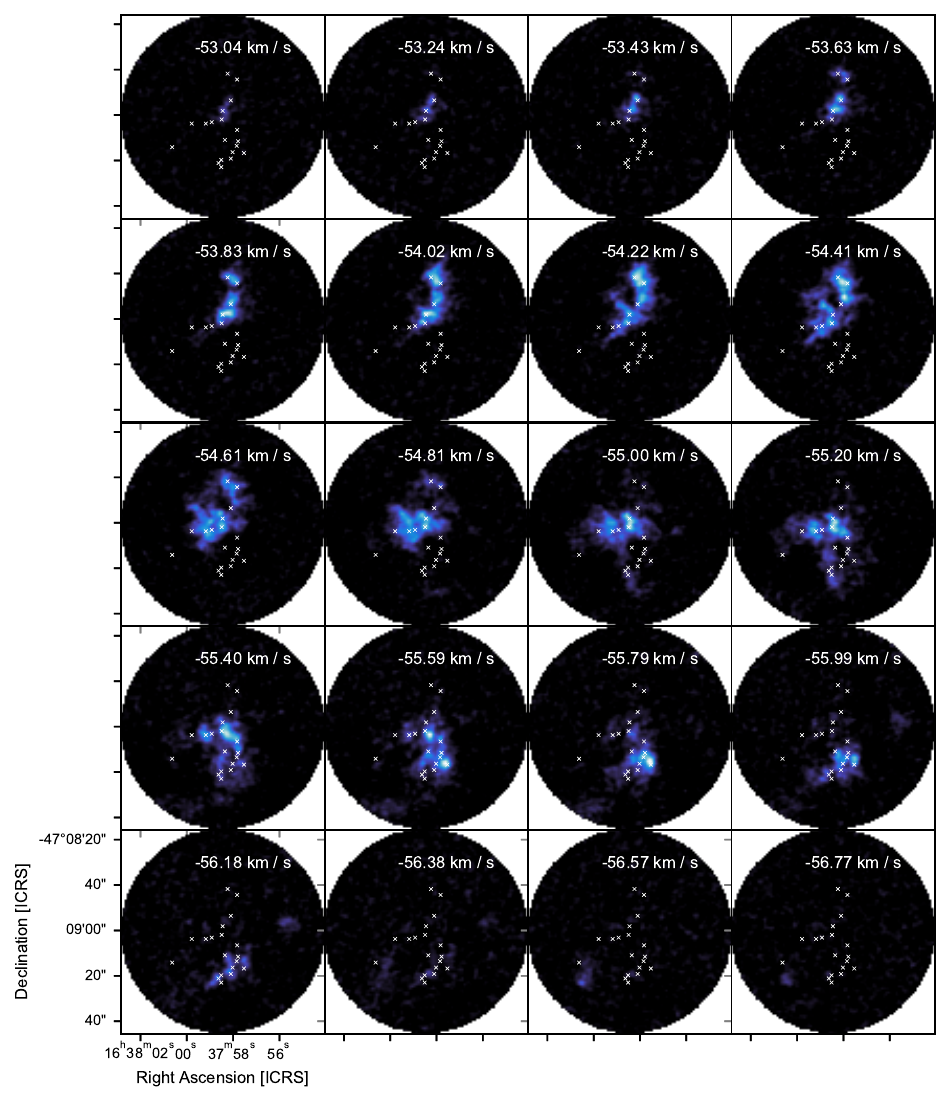}
    \caption{Channel map of N$_2$H$^+$ ($J=1-0$, $F_1, F=0,1-1,2$). The white crosses are plotted at the peaks in the continuum of the identified cores in \citet{Morii23}. }
    \label{fig:channel_map}
\end{figure*}
We confirm these features also on the channel map (Figure~\ref{fig:channel_map}) where the core positions identified in \citet{Morii23} are overlaid. 
Overall, as the velocity decreases, emission shifts from north to south, and the N$_2$H$^+$ emission and core positions seem to match well as seen in the moment maps. 
Upon careful inspection of the channel map, we find a different distribution between the red and blue shifted sides with respect to the systemic velocity of the clump (--54.6 km\,s$^{-1}$). 
On the red-shifted side, the higher velocity component first appears near the center and extends to the north. 
This may be a sign of the accelerated gas toward the central cores, which will be addressed later (sec. \ref{sec:pv}). 
However, the blue-shifted component extends from the lowest velocities in the central parts to the highest velocities in the outer, southern regions. 
From the channel map, we find that the multicomponents are localized near the center and not spread out as seen in more evolved clusters \citep[e.g.,][]{Henshaw14,Rigby24, almaimf_n2hp_24_arxiv}. 

\begin{figure*}
    \centering
    \gridline{\fig{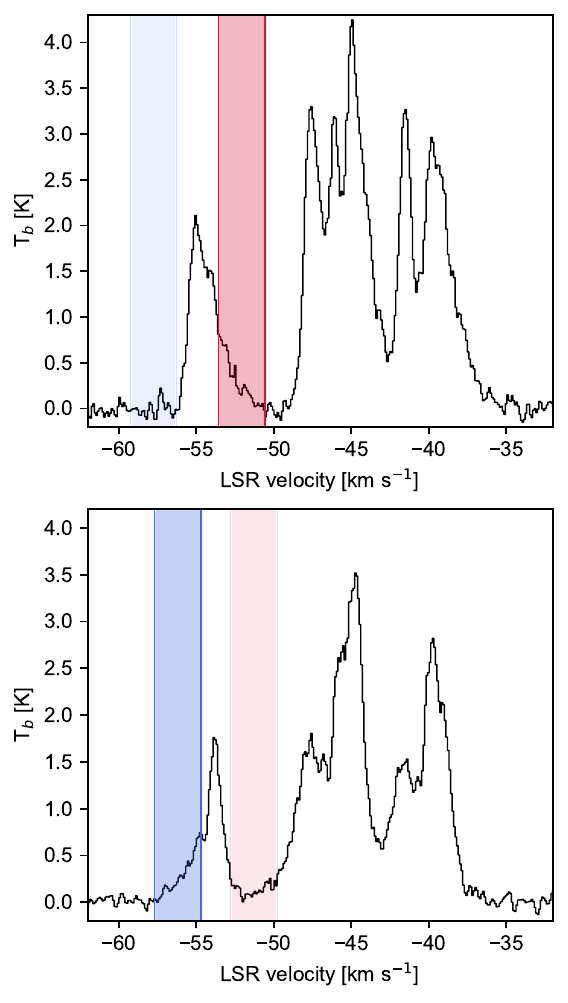}{0.24\textwidth}{(a) N$_2$H$^+$ spectra } 
              \fig{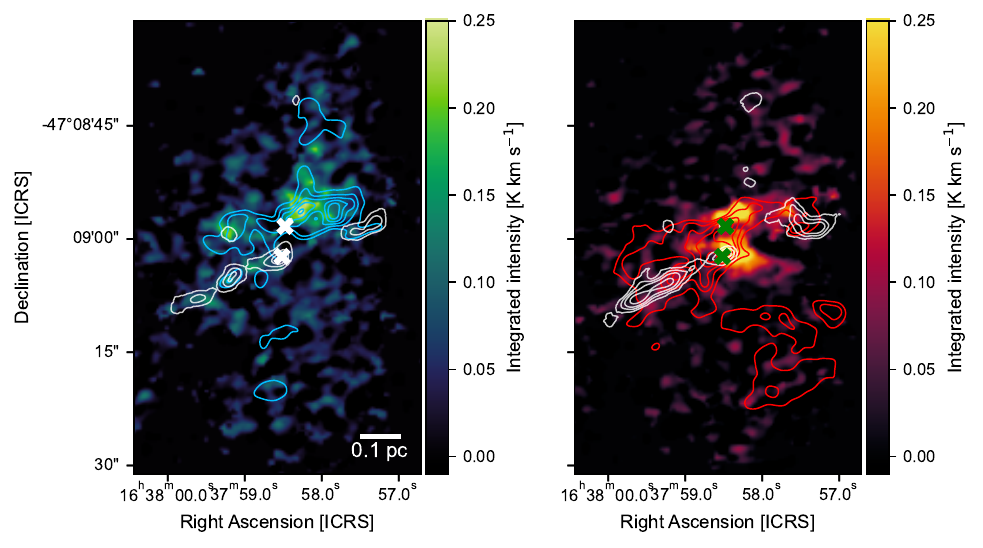}{0.75\textwidth}{(b) Tail maps}
             }
    \caption{(a) Example of the N$_2$H$^+$ ($J = 1-0$) spectra showing tail components (a reference frequency of 93.17626 GHz). The velocity range used to make the tail maps is highlighted as the shaded zone.  (b) The integrated intensity maps of (left) the blue-shifted and (right) the red-shifted wing components in the raster, CO ($J = 2-1$) in red and light-blue contours and SiO ($J=5-4$) in white contours. The integration velocity range for CO and SiO is almost same as the shaded range in panel (a) as $v_{\rm lsr}= -58.8$ to -55.8 km\,s$^{-1}$ and $v_{\rm lsr}= -53.6$ to -50.6 km\,s$^{-1}$.  Positions of ALMA1 and ALMA2 are plotted as white/green symbols. } 
    \label{fig:tailmap}
\end{figure*}

However, the line profile of the whole cube is complex. 
Figure~\ref{fig:tailmap} (a) shows examples of the spectra in small regions near ALMA1 and ALMA2.   
Not only are the expected hyperfine components detected, but additional components are also seen.  
Even the optically thinner, isolated hyperfine component has a wing or tail, resembling those often observed in outflow tracers, as emphasized in the shaded regions in each panel. 
To understand what such wing components are, we made the integrated intensity maps of the wing/tail components using the peak velocity map ($v_{\rm peak}$) and the velocity dispersion (mom 2) map. 
We integrate the velocity range from $v_{\rm peak} \pm HWHM $ to $v_{\rm peak }\pm HWHM \pm 3\,{\rm km\,s^{-1}}$, where HWHM is the half-width, half-maximum estimated from the intensity-weighted average velocity dispersion. 
The plus and minus sign correspond to the red and blue shifted components, respectively. 
Near the center of the clump, the HWHM is $\sim$0.6--1 km\,s$^{-1}$ and the integrated velocity range is $|v_{\rm lsr} - v_{\rm peak}| \sim 0.8-3.8$\,km\,s$^{-1}$.  
The 3 km\,s$^{-1}$ interval was determined to maximize the range but to avoid the neighboring hyperfine components, especially for the red-shifted side.  
Examples of the velocity range are plotted in the spectra as red and blue shades.  
The produced maps are shown in Figure~\ref{fig:tailmap} (b). 
Both blue-shifted and red-shifted tail components are bright around the center of the clump, where outflows are detected. 
In the figure, CO and SiO emission are overlaid as red/light-blue and white contours, respectively. 
The integration velocity range for CO and SiO is the same as for the N$_2$H$^+$ wing components, and is $v_{\rm lsr}= -58.8$ to -55.8 km\,s$^{-1}$ and $v_{\rm lsr}= -53.6$ to -50.6 km\,s$^{-1}$, for blue-shifted and red-shifted components, respectively.  
The positions of ALMA1 and ALMA2 are plotted as x symbols. 
For the blueshifted component, N$_2$H$^+$ is bright inside CO contours which traces outflow from ALMA1.  
The redshifted N$_2$H$^+$ tail components exist near ALMA1 and ALMA2 and partially overlapped with CO and SiO emission, likely corresponding to the outflow surface. 
N$_2$H$^+$ is typically known to be a high-density tracer of cold quiescent gas \citep{Sanhueza12,Sanhueza13} but in our observations it has high-velocity components with a spatial distribution similar to that of the outflow tracers. 
In the particular case of G337, the N$_2$H$^+$ ($J = 1-0$) transition is likely to trace the dense gas entrained by protostellar outflows. 
In low-mass star-forming regions, similar cases have been reported in \citet{Tobin11} and \citet{Bjerkeli16}. 
This can be a first report of the N$_2$H$^+$ ($J = 1-0$) tracing entrained gas in high-mass star-forming regions. 

\subsection{HNC ($J = 3-2$) and HCO$^+$ ($J = 3-2$)} 
\begin{figure*}
\begin{tabular}{cc}
    \begin{minipage}[b]{0.8\linewidth}
        \centering
        \includegraphics[keepaspectratio]{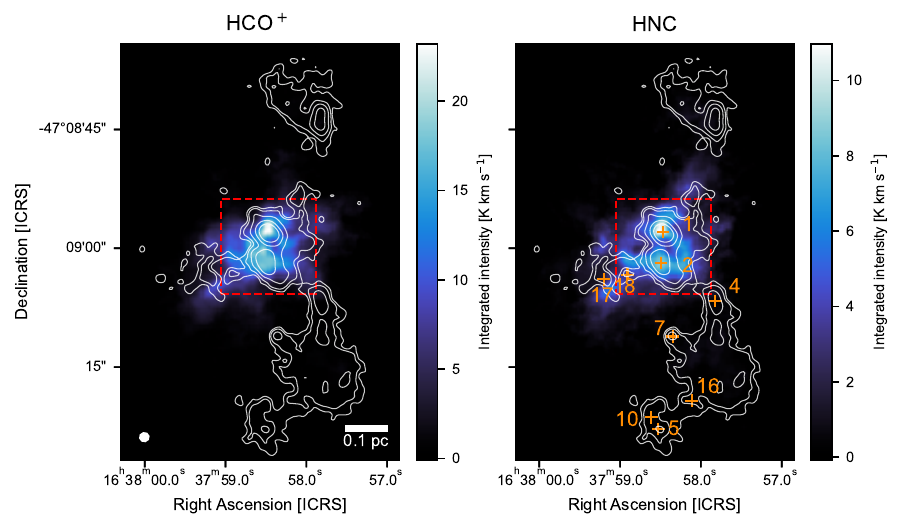}
        \caption{The mom0 maps of HCO$^+$ ($J = 3-2$) and HNC ($J = 3-2$), overlaid with continuum emission as white contours. The contour levels are the same as in Figure~\ref{fig:n2hp_mom}. }
        \label{fig:hnc_mom0}
      \end{minipage} \\
      \begin{minipage}[b]{0.6\linewidth}
        \centering
        \includegraphics[keepaspectratio]{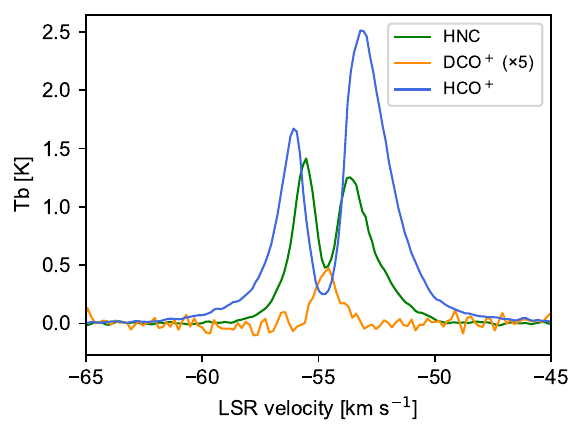}
        \caption{Line spectra of HCO$^+$ (blue), HNC (green), and DCO$^+$ (orange), averaged within the red square in the left panel (a). The orange crosses and the numbers represent the continuum peak positions of the cores and their id, which are discussed in Section~\ref{sec:hill5}. }
        \label{fig:hnc_ave}
      \end{minipage}
\end{tabular}
\end{figure*}
Figure~\ref{fig:hnc_mom0} shows the integrated intensity maps of HCO$^+$ and HNC ($J = 3-2$). 
Both emission lines show similar spatial distributions and are bright near the central cores, consistent with their higher critical density ($\gtrsim$10$^6$ cm$^{-3}$ at 10\,K) compared to that of N$_2$H$^+$ ($\sim$6$\times$10$^4$\,cm$^{-3}$ at 10 K), although the extended emission in the north-west direction is similar to N$_2$H$^+$. 
Their spectra averaged within the red square in the figure are displayed in Figure~\ref{fig:hnc_ave}. 
Together with HCO$^+$ (blue) and HNC (green), we overplotted the spectra of DCO$^+$ ($J = 3-2$) as an optically thin tracer. 
DCO$^+$ has a peak around $v_{\rm lsr}=-54.5$ km\,s$^{-1}$, and both HCO$^+$ and HNC show dips at this velocity, indicating that the dip is produced by self-absorption. 
Although the dip of HCO$^+$ is slightly shifted, the difference is just one or two spectral channels. 
There is a clear difference between the line profiles of HCO$^+$ and HNC in peak velocity, red is bright in HCO$^+$, but blue is brighter in HNC. 
Furthermore, the line profile of HCO$^+$ is broader than that of HNC with high-velocity tails greater than 5\,km\,s$^{-1}$. 
These high-velocity components are detected in areas where the CO outflow or the SiO jet are identified.
This implies that HCO$^+$ is more contaminated by the gas entrained by the outflow and/or jet. 

\section{Analysis}
\label{sec:analysis}
\subsection{Velocity gradients in PV diagram}  \label{sec:pv} 
\begin{figure}
    \centering
    \includegraphics[width= 8.6cm]{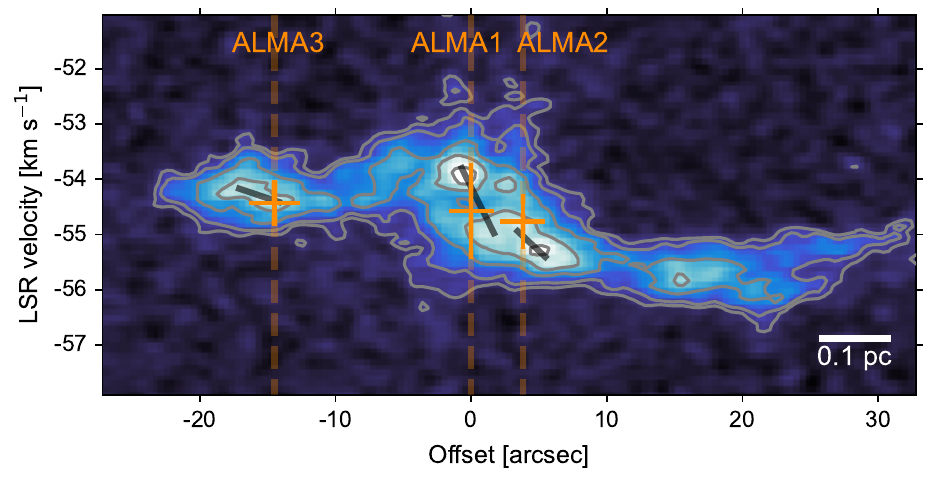}
    \caption{The position-velocity diagram of N$_2$H$^+$ ($J=1-0$, $F_1, F=0,1-1,2$) along the declination with a width of 10 arcsec, centered on the continuum peak position of ALMA1. The contour levels are 3, 5, 7, 10, 15, 20 $\sigma$, where 1 $\sigma$ = 2.5 mJy beam$^{-1}$. The vertical orange lines are plotted at the continuum peak positions of ALMA3, ALMA1 and ALMA2, from left to right, respectively. The orange plus symbols indicate core positions (x-value), systemic velocities (y-value), core size (x-range), and FWHM (y-range). The black lines show the velocity gradient discussed in the main text.  
    The scale bar is shown on the bottom right. }
    \label{fig:pv}
\end{figure}
The position-velocity (PV) diagram is one of the useful tools for characterizing processes such as infall, outflow, or rotation. 
We produced the PV diagram along the declination with a 10\arcsec width to cover most of the continuum emission (black shaded region in Figure~\ref{fig:n2hp_mom}). 
The PV diagram is presented in Figure~\ref{fig:pv}. 
The center in the position axis is taken at the continuum peak of ALMA1, and the negative offsets correspond to the north part. 
Some core positions (ALMA1, ALMA2, and ALMA3 in \citet{Morii23}) are marked as orange lines. 
The overall velocity gradient from north to south and the increase of velocity close to the continuum source found in the red-shifted side, seen in the moment maps or the channel maps, are again confirmed from the PV diagram. 
Around ALMA1 and ALMA2, we see a complex, sharp transition, which looks similar to the picture reported in a massive cluster-forming clump, NGC 2264-C \citep{Peretto06}. 
The curvy structures from the left to the center can be interpreted as the accretion flow to the central subclump. 
However, the right part looks more flat from the south to ALMA2. 
These different features seen in the northern and southern regions may come from the different orientations of the clump structure. 

In addition, focusing on the localized bright sections of the PV diagram, we find smaller-scale ($\sim$0.04 pc) gradients around three cores (offset of -15\arcsec, 0\arcsec, and 4\arcsec). 
The orange plus symbols in Figure~\ref{fig:pv} indicate the positions of the cores and their systemic velocity by the crossing point, and their size and FWHM by the size of the bars. 
Pairs of bright peaks exist around the plus symbols and are highlighted as black segments in Figure~\ref{fig:pv}. 
Such velocity gradients can be produced by infalling motions or solid-body rotation. 
The velocity gradients measured in the position-velocity diagram are $\nabla v_{\rm obs}= 25.6$, 11.8, and 4.25 km\,s$^{-1}$\,pc$^{-1}$ for ALMA1, ALMA2, and ALMA3, respectively. 
These values are all larger than the gradients measured in Cygnus-X at 0.1 pc scale \citep[1.2--4.2 km\,s$^{-1}$\,pc$^{-1}$][]{Csengeri11} and in low-mass cores \citep[0.3--4 km\,s$^{-1}$\,pc$^{-1}$][]{Goodman93} where they assumed the velocity gradient due to rotation. 
The very roughly estimated specific angular momentum assuming the velocity gradient caused by the rotation was 0.01, 0.005, and 0.003 km\,s$^{-1}$\,pc$^{-1}$ for ALMA1, ALMA2, and ALMA3, respectively. 
All are greater than the prediction of 10$^{-3}$ km\,s$^{-1}$\,pc$^{-1}$ extrapolated from the correlation derived in \citet{Goodman93} especially for ALMA1. 
Furthermore, we did not find rotation features by checking optically thin tracers of warm gas (CH$_3$OH and HC$_3$N). 
We therefore discard solid-body rotation and conclude that the most likely explanation is infalling motions. 
We speculate that observations at higher angular resolution would reveal signs of acceleration near the protostars, as expected from an infalling envelope \citep[e.g.,][]{Sanhueza21, Ahmadi23, Olguin23}. 
We estimate the crossing time of these flows by $t_ {\rm cross} = 1/\nabla v_{\rm obs}$ to be 38, 83, and 230 kyr, respectively. 
They are comparable to the free-fall time of cores (10--70 kyr), implying that the observed gradients come from infall motions, and the crossing time provides a good estimate for an infall timescale.  
If the observed cores have formed within these crossing times, then the typical infall rates are $\dot{M}_{\rm obs} = M_{\rm core} / t_{\rm cross}$. 
The mass infall rate after correcting for the effect of the angle of inclination $i$ is given by $\dot{M} = \dot{M}_{\rm obs} {\rm tan}\,i$. 
Assuming a moderate inclination angle of $i = 45\deg$ following \citet{Chen19}, $\dot{M}$ is
(2.7, 0.49, 0.23) $\times$10$^{-4}\,M_\odot$\,yr$^{-1}$, respectively. 

\begin{figure}
    \centering
    \includegraphics[width=8cm]{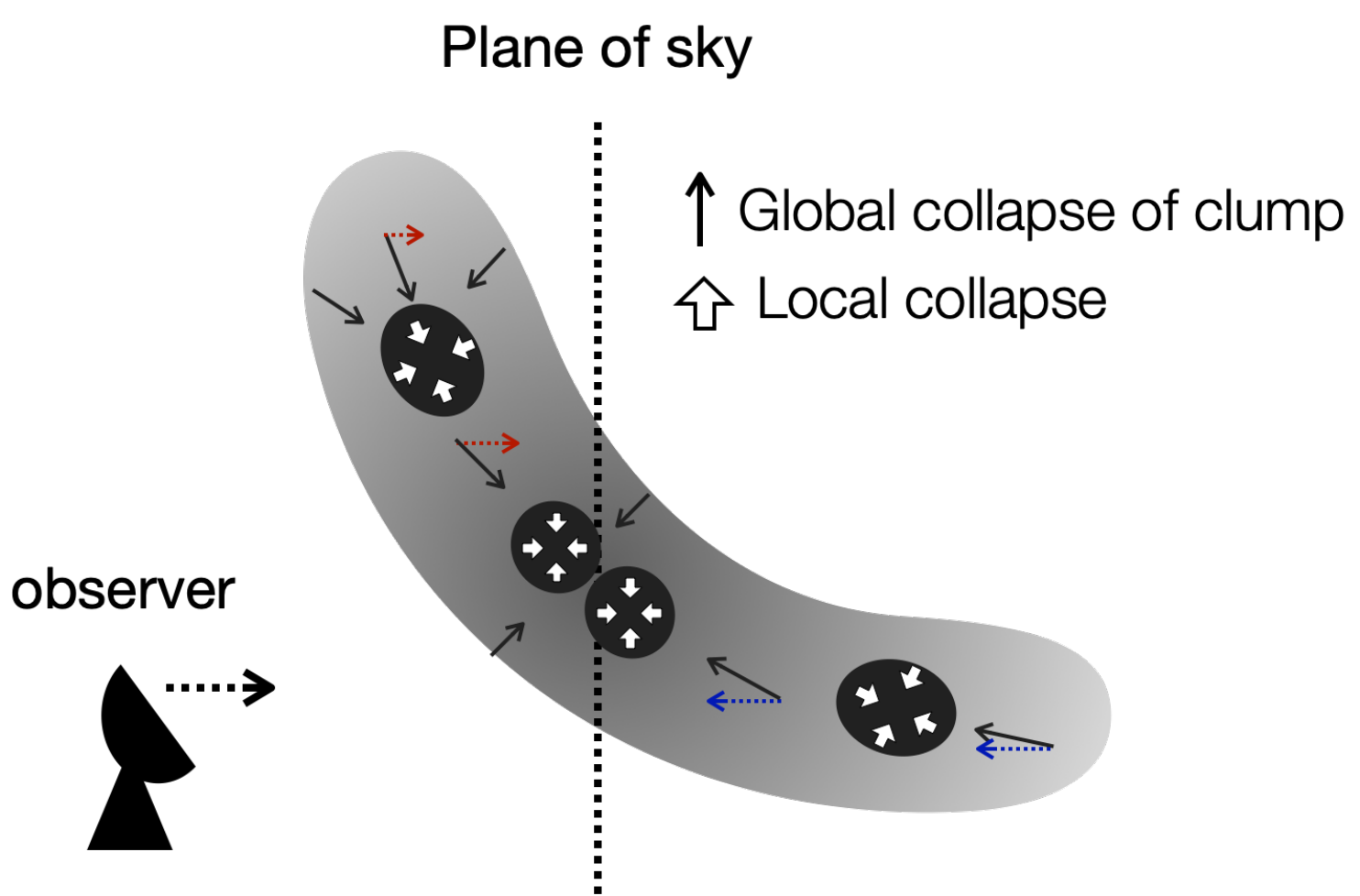}
    \caption{Schematic picture of the gas dynamics in G337}
    \label{fig:schematic_fig}
\end{figure}

Figure~\ref{fig:schematic_fig} shows a schematic picture of our proposed scenario that reproduces the kinematic structures observed in N$_2$H$^+$. 
Our analysis with N$_2$H$^+$ revealed the clump-scale velocity gradient with different features in the northern and southern regions of the IRDC. 
Furthermore, we also find smaller scale gradients implying localized infall around cores. 
Observed features can be explained if the clump has a different inclination angle against the plane of sky in the northern and southern regions, and gas flow exists toward the central part due to the gravitational collapse of the clump, as the figure shows.  
Another possible idea is the convergence of the two different velocity components ($v_{\rm lsr} \sim$-54.5 km\,s$^{-1}$ and -55.5 km\,s$^{-1}$) at the center of the clump. 
The velocity difference between the two components are $\sim$1 km\,s$^{-1}$ from the PV diagram, similar to the velocity dispersion of the clump \citep[0.8 km\,s$^{-1}$;][]{Sanhueza19}. 

\subsection{Blue-asymmetry profile} \label{sec:hill5}  
Optically thick dense gas tracers, such as HCO$^+$ and HNC line emission, have been used to search for infall motions revealed by the so-called blue asymmetry profile, characterized by a double-peaked line profile with brighter blue emission \citep[e.g.,][]{Fuller05, Contreras18, Jin-Jin21}. 
This profile has been interpreted as a sign of infall within a collapsing core having a centrally increasing density profile and two points along a line of sight with the same velocity \citep[][and his Figure~5]{Evans99}. We note that the identification of such a profile depends on the viewing angle in the case of a non-spherical core, or a core embedded in a filament \citep[e.g.,][]{Smith12}. 
Our setup includes HCO$^+$ ($J = 3-2$) and HNC ($J= 3-2$) as candidates for line emission tracing infall motions.  
These $J=3-2$ transitions are expected to be suitable lines to study infall motions in high-mass star formation, as determined from both numerical simulations \citep[e.g.,][]{Chira14} and observations \citep[e.g.,][]{WuEvans10, Jin-Jin21}. 
Figure~\ref{fig:hnc_ave} displays the observed line emission averaged within the red square in Figure~\ref{fig:hnc_mom0}. 
As discussed in the previous section, the HNC is less affected by outflow emission and shows a blue asymmetry profile, implying that the central part of the massive clump (or subclump) is collapsing. This is consistent with the virial parameter of this sub-clump ($\alpha \sim$0.5). 
In the following, we focus our analysis on the HNC line data because it appears to be less affected by outflow emission.  
To characterize the infall motion, we modeled the HNC line emission from the cores. 
We carefully checked the line profile averaged inside the ellipse identified by astrodendro \citep{Morii23}{}, and found that 9 of 17 dust cores (53\%) have HNC emission with a blue-asymmetry profile. 
These nine cores are highlighted in Figure~\ref{fig:hnc_mom0} as orange crosses. 
Among the nine cores showing the blue asymmetry profile, two cores contain high-velocity wings in the red-shifted component. The high-velocity component was removed before modeling the line by multiple-Gaussian fitting (see Appendix~\ref{sec:Appendix_outflowextraction} for details). 

\begin{figure*}
    \centering
    \includegraphics[width=18cm]{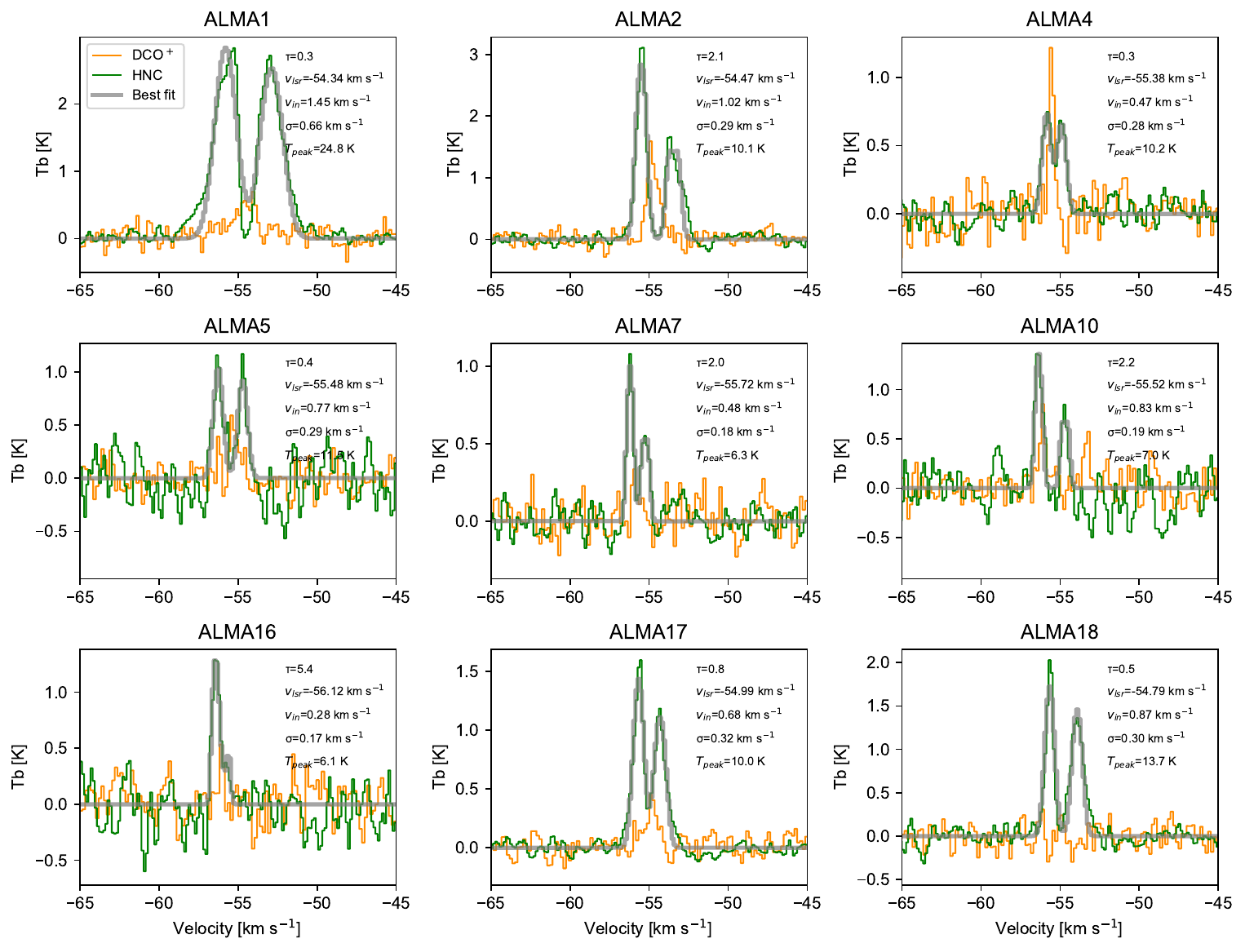}
    \caption{Line spectra of HNC (green) and DCO$^+$ (orange) of nine cores averaged inside cores. The thick gray lines represent the results of the Hill5 model fit. The best-fit parameters are shown on the right of each panel. }
    \label{fig:HNC_fitting_all}
\end{figure*}
The line profiles of the HNC and DCO$^+$ lines averaged within the nine cores are shown in Figure~\ref{fig:HNC_fitting_all}, in green and orange, respectively. 
These are extracted from the primary beam-corrected cubes. 
Some cores such as ALMA2, 7, 17, and 18 show the blue-brighter profile with a dip around optically thin line peak velocity, while others such as ALMA1, ALMA4, and ALMA5 have a similar brightness in red and blue shifted peaks. 
To estimate the infall velocity from these spectra, we applied the Hill5 model \citep{De_Vries05}.  
Hill5 model is a simple radiative transfer model that can reproduce the observed spectral asymmetries, which are expected to arise in a collapsing core. 
In this model, the core has a peak excitation temperature of $T_{\rm peak}$ at the center, and it has an excitation temperature of $T_0$ =2.3\,K at the edges of the core. 
Thus, the core is modeled as a two-layer slab, where the excitation temperatures increase linearly up to a peak temperature at the boundary between the two regions ($T_{\rm peak}$), and then decrease linearly back to the initial temperature ($T_0$). 
For this model, there are five free parameters to fit; 1. the optical depth of the line ($\tau$), 2. the systemic velocity of the structure ($v_{\rm lsr}$), 3. the infall velocity of the gas in the core ($v_{\rm in}$), 4. the velocity dispersion of the molecular line ($\sigma$), and 5. the peak excitation temperature ($T_{\rm peak}$). 
This model may underestimate the infall velocity in some cases. 
However, the reliability of the model improves when the line profile exhibits separate blue- and red-shifted components \citep{De_Vries05}.  

We applied the Hill5 model to the averaged spectra. 
The actual fit was conducted using the affine-invariant Markov Chain Monte Carlo (MCMC) algorithm implemented in the emcee Python package \citep{Foreman-Mackey-emcee} to explore the parameter space. 
We run 100 walkers for 15,000 steps.  
For the parameter space, we used $\tau$ ranging from 0.1 to 30, a $v_{\rm lsr}$ between -57 and -54 km\,s$^{-1}$, $v_{\rm in}$ between 0.1 and 4 km\,s$^{-1}$, $\sigma$ between 0.1 and 1.5 km\,s$^{-1}$, and $T_{\rm peak}$ between 2 and 100 K. 
The best-fit parameters and the spectra are shown in Figure~\ref{fig:HNC_fitting_all}.   
The estimated infall velocity ranges from 0.28 to 1.45 km\,s$^{-1}$, and the velocity dispersion $\sigma$ ranges from 0.2 to 0.6 km\,s$^{-1}$, which is comparable to the estimated velocity dispersion from the DCO$^+$ line in \citet{Li23}. 
ALMA1 has the largest $v_{\rm in}$ and $\sigma$. 

The mass infall rate of the core was calculated using $\dot{M} = 4 \pi\,R^2\,\rho\,v_{\rm in}$. 
For example, ALMA1 has the volume density of $\rho=4.21\times$10$^{-17}$ g\,cm$^{-3}$ and the radius of the core of $R=3250$ au \citep{Morii23}. 
With $v_{\rm in}$ estimated from the fitting, we derived a mass infall rate of 2.66 $\times$10$^{-3} \,M_\odot$\,yr$^{-1}$. 
This value may be underestimated if the model underestimates the infall velocity.  
The estimated infall velocity and rate are comparable to the one estimated in an intermediate-mass prestellar-core candidate embedded in another 70 $\mu$m-dark clump \citep[1.96$\times$10$^{-3} M_\odot$\,yr$^{-1}$ by ][]{Contreras18}. 
The infall rate calculated for the other cores is summarized in Table~\ref{tab:mcmc}, generally one order of magnitude smaller than that of ALMA1 except for ALMA2.

\begin{deluxetable}{lccccc}
\label{tab:mcmc}
\tabletypesize{\footnotesize}
\tablecaption{Core properties and derived infall parameters}
\tablewidth{0pt}
\tablehead{
\colhead{Core name}  & \colhead{$R$} & \colhead{$M$} & \colhead{$\rho$} & \colhead{$v_{\rm in}$} & \colhead{$\dot{M}_{\rm infall}$} \\ 
\colhead{} & \colhead{au} & \colhead{$M_\odot$} & \colhead{10$^{-17}$ g\,cm$^{-3}$} & \colhead{km\,s$^{-1}$} & \colhead{10$^{-4}$\, $M_\odot$} }
\startdata
ALMA1 & 3250  & 10.20 & 4.21 & 1.45$\pm$0.01 & 28.9 \\
ALMA2 & 3380  & 4.11 & 1.51 & 1.02$\pm$0.01 & 7.8 \\
ALMA4 & 2680  & 1.21 & 0.89 & 0.47$\pm$0.02 & 1.3 \\
ALMA6 & 1930  & 0.96 & 1.87 & 0.77$\pm$0.02 & 2.4 \\
ALMA7 & 2840  & 1.15 & 0.71 & 0.48$\pm$0.02 & 1.2 \\
ALMA10 & 1550  & 0.42 & 1.60 & 0.83$\pm$0.01 & 1.4 \\
ALMA16 & 1280  & 0.20 & 1.38 & 0.28$\pm$0.03 & 0.3 \\
ALMA17 & 5960  & 1.37 & 0.09 & 0.68$\pm$0.01 & 1.0 \\
ALMA18 & 1960  & 0.24 & 0.45 & 0.87$\pm$0.01 & 0.7 \\
\enddata
\end{deluxetable}

\section{Discussion} \label{sec:discussion}
\subsection{Infall velocity and rate in G337} 
\begin{figure*}
    \centering
    \includegraphics[width=18cm]{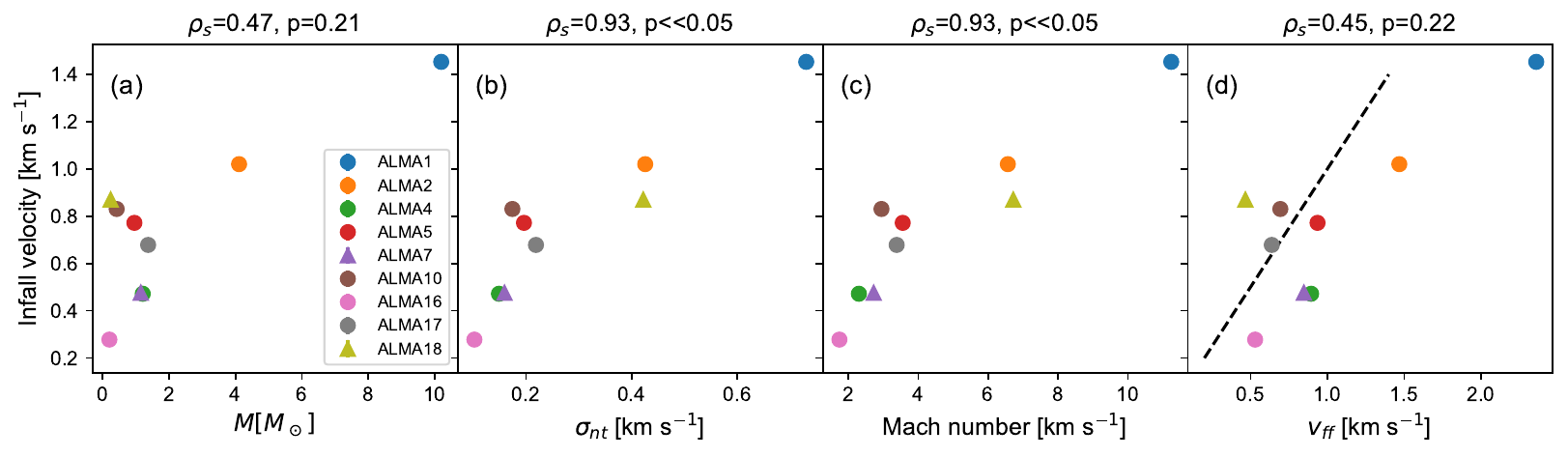}
    \caption{Scatter plots of infall velocity with (a) core mass, (b) nonthermal velocity dispersion, (c) Mach number, and (d) free-fall time. The velocity width was derived from DCO$^+$, but for two cores without DCO$^+$ detection, we measured them from N$_2$H$^+$ and highlighted them as triangle shapes. On the top of each panel the Spearman's rank correlation coefficient ($\rho_s$) and the p-value are shown. }
    \label{fig:vin_corephy}
\end{figure*} 
We succeeded in estimating infall velocities from intermediate-mass and low-mass cores embedded in a 70 $\mu$m-dark massive clump.  
In this section, we investigate the correlation of these velocities with the core physical properties. 
Figure~\ref{fig:vin_corephy} shows the scatter plots of the infall velocity with (a) core mass, (b) nonthermal velocity dispersion, (c) Mach number and (d) free-fall time of the cores. 
As shown in panel (a), two intermediate cores have a higher infall velocity than low-mass cores, resulting in a moderate correlation between $v_{\rm in}$ and the mass of the core, although between low-mass cores ($M <$1\,$M_\odot$) the correlation is weak. 
The Spearman’s rank correlation coefficients ($\rho_s$) between $v_{\rm in}$ and the core mass are 0.47, and the p-value is 0.2. 
We find a stronger correlation of the infall velocity with (b) the nonthermal velocity dispersion and with (c) the Mach number, the last two quantities derived from DCO$^+$ (or N$_2$H$^+$ for two cores). 
With an $r_s >$0.9 and a p-value much lower than 0.05, we conclude that the nonthermal component derived from optically thin tracers is contaminated by infall motions, and it is not purely turbulence. 
We suggest that this contamination has to be significant to find such strong correlations.  
This is consistent with the conclusion made by \citet{Traficante18} from a clump-scale study, based on the correlation between the gravitational acceleration term and the kinetic term. 
Panel (d) compares $v_{\rm in}$ with free-fall velocity cores, $v_{ff} = - \sqrt{2GM_{\rm core}/R_{\rm core}}$, where $G$ is a gravitational constant, $M_{\rm core}$ is the mass of the core and $R_{\rm core}$ is the radius of the core. We used $M_{\rm core}$ and $R_{\rm core}$ from \citep{Morii23}. 
The dashed line corresponds to $v_{\rm in}$ = $v_{\rm ff}$. Six cores show $v_{\rm in} < v_{\rm ff}$, but the difference between these two velocities is within a factor of two for all cores, which implies that the observed infall velocity is comparable to the free-fall velocity. 

\begin{figure}
    \centering
    \includegraphics[width=8.5cm]{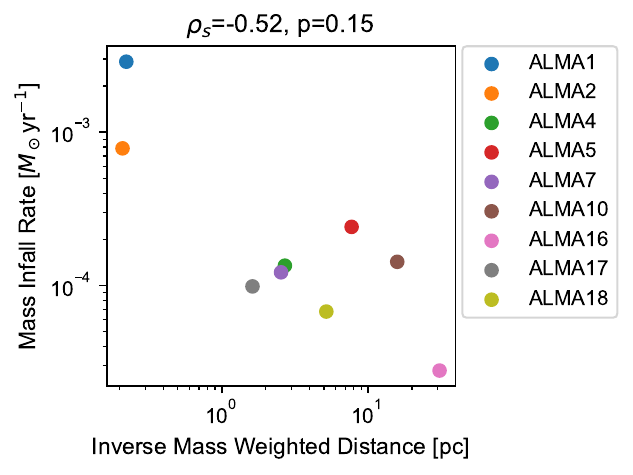}
    \caption{A scatter plot of mass infall rate with inverse mass-weighted distance, $\tilde{d} (M_i)$. As same with Figure~\ref{fig:vin_corephy}, the Spearman's rank correlation coefficient ($\rho_s$) and the p-value are shown on the top.}
    \label{fig:Mdot_Mdist}
\end{figure}

Clump-fed models predict that the higher infall velocity or mass infall rate will occur near the cluster centers, making these cores the most massive ones that will likely form high-mass stars. 
In G337, two intermediate-mass cores are located near the cluster center. To inspect this prediction, we define a new parameter that takes into account the distance from the cluster center and their masses, the inverse mass-weighted distance, as
\begin{equation}
    \tilde{d} (M_i) = (x_i-x_0) / (M_i / \Sigma M_j),
\end{equation}
where $x_i$ and $M_i$ are the position and mass of the core $i$, and $x_0$ is the position of the cluster center.  
We assume the gravity potential center as the center of the cluster, calculated from the continuum emission map assuming a constant temperature. The position $x_0$ then corresponds to R.A. (J2000.0) = $16^h 37^m 58 \fs 613$ and Dec (J2000.0) = $-47^\circ 09^\prime 03\farcs 830$. 
The correlation plot of the mass infall rate and the inverse mass-weighted distance is shown in Figure~\ref{fig:Mdot_Mdist}. 
As expected from panel (a) in Figure~\ref{fig:vin_corephy}, the two intermediate mass cores have higher mass infall rates with smaller $\tilde{d} (M_i)$. 
The Spearman's rank correlation coefficient ($\rho_s$) is -0.52, and the p-value is 0.15, indicating strong anti-correlation. 
This is consistent with clump-fed models.  

\begin{figure}
    \centering
    \includegraphics[width=7cm]{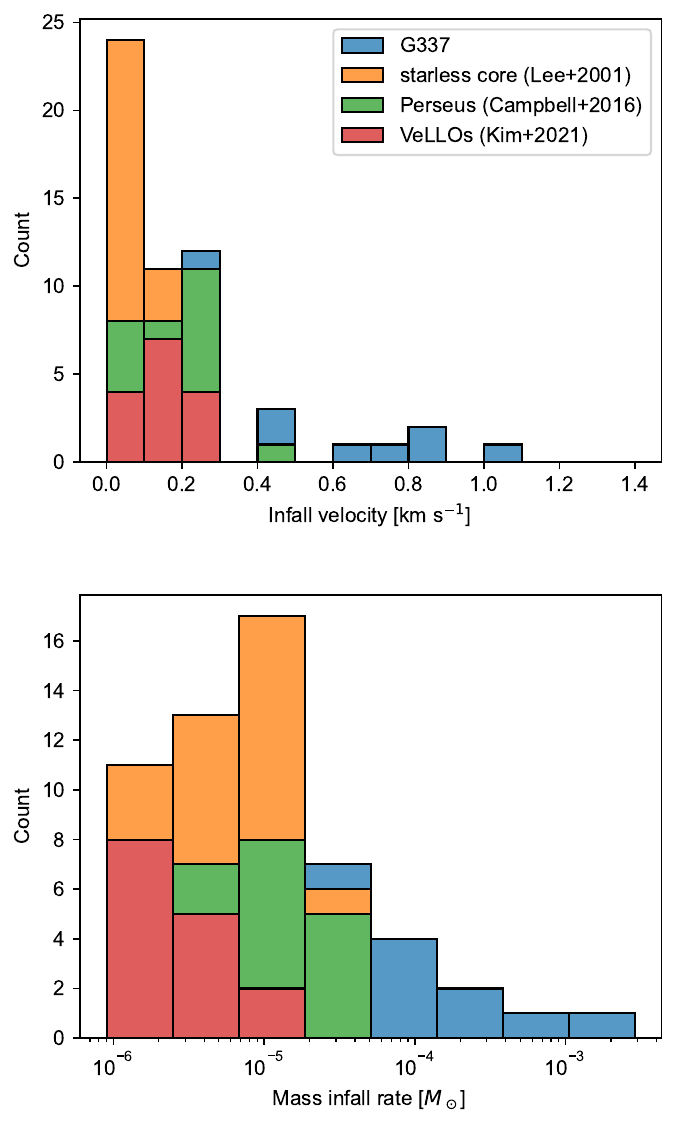}
    \caption{Histogram of infall velocity and infall rate compared with low-mass star forming regions.}
    \label{fig:Vin_Mdot_hist}
\end{figure} 
To characterize the derived infall properties of cores in the G337 region, we compared the infall velocity, mass infall rate, and core mass with those in low-mass star forming regions. 
The data of cores in nearby low-mass star-forming regions is from a starless core catalog \citep{Lee01}, Perseus cores \citep{Campbell16}, and VeLLOs catalog \citep{Kim21}, where the mass distribution and the angular resolution are similar to our sample. 
The infall velocity and infall rate are all estimated by the Hill5 model, while the molecular line used, and observational setups (interferometer or single-disk telescopes) are different. 
Figure~\ref{fig:Vin_Mdot_hist} shows the comparison of the infall velocity and the mass infall rate. 
Compared with cores in low-mass star-forming regions, the observed infall velocity in G337 is clearly several times higher in $v_{\rm in}$ and one order of magnitude higher in $\dot{M}$. 
We checked that the mass distribution, line widths of optically thin tracers (e.g., DCO$^+$ in our case and N$_2$H$^+$ in low-mass star-forming regions) and temperature are similar within the sample, but our cores are a few times smaller and more than one order of magnitude denser. 
We conclude that the environment plays an important role in the star formation process, especially in the formation of high-mass stars. 
Low-mass cores with similar properties have significantly different infall speeds and mass infall rates when they are in lower density, low-mass star-forming regions with respect to denser high-mass star-forming regions. 
The higher mass infall rate onto cores in high-mass star-forming regions most likely contributes to the core growth, allowing the accumulation of sufficient mass over time to eventually form high-mass stars from cores that were initially of intermediate mass.

\subsection{Are cores growing to be massive?} 
Recent observational studies in the early stages of high-mass star formation revealed the lack of high-mass prestellar cores and a large population of gravitationally bound low- to intermediate-mass cores \citep[e.g.,][]{Sanhueza17, Svoboda19, Morii23}. 
These studies support the clump-fed scenario and expect core growth through the gas feeding from the vicinity toward cores. 
Indeed, large surveys using ALMA across various evolutionary stages imply the mass growth of cores as the cores evolve (ASHES; \citealt{Li23}, ALMA-IMF; \citealt{Pouteau23}, SQUALO \citealt{Traficante23}, ATOMS; \citealt{Xu23}, QUARKS; \citealt{Xu24}, and ALMAGAL; \citealt{Wells24}).  
Some previous case studies also support this scenario by the detection of velocity gradients or the infall profile around the cores and the derived high mass-inflow rates of the order of 10$^{-3} M_\odot$\,yr$^{-1}$ \citep[e.g.,][]{Peretto14, Contreras18, Redaelli22}. 

Our analysis in G337 revealed signs of infall from the position-velocity diagram and more directly from the infall profile, implying that infall motion around cores actively occurs even in the early stages of cluster formation. 
The derived mass infall rate is on the order of several times 10$^{-4}-10^{-3} M_\odot$\,yr$^{-1}$ from both analysis, although the mass infall rate derived from the velocity gradient is generally smaller.   
These values are quite higher compare to  the infall rates in low-mass star-forming regions, as discussed in the previous section, and also than the outflow rate of the cores in this region \citep[$\sim$10$^{-6}\,M_\odot$\,yr$^{-1}$;][]{Li20}. 
Additionally, the timescale derived from the velocity gradient was comparable to the free-fall time and the infall velocity estimated from the line profile was similar to the free-fall time.  
Assuming that these high infall rates continue for a free-fall time, the most massive core can gain an additional 30\,$M_\odot$. 
In fact, the central subclump, which hosts two cores, has a mass of $>$100\,$M_\odot$, high enough to feed the cores. 

\section{Conclusions}
\label{sec:conclusion} 
We have presented the ALMA study of gas dynamics in a 70 $\mu$m dark massive clump, G337.541--00.082, using dense gas tracers such as N$_2$H$^+$ and HNC to reveal the dynamical infall motion around cores expected to achieve the formation of high-mass stars. 
We have obtained the following conclusions: 
\begin{enumerate}
    \item N$_2$H$^+$ ($J = 1-0$) line is detected across the whole clump, and well overlapped with continuum emission as well as cores. It shows a clump-scale velocity gradient and has a larger velocity dispersion around the central intermediate-mass protostellar cores. Comparison of high-speed tail components with CO and SiO outflow/jet emission implies that N$_2$H$^+$ can also trace outflow lobes. 
    
    \item HNC ($J = 3-2$) and HCO$^+$ ($J = 3-2$) lines are detected around the densest part of the clump and exhibit similar spatial distributions. However, after inspecting the line profiles, HCO$^+$ is more severely affected by outflows with high-velocity wings/tails than HNC. 

    \item The position-velocity diagram of N$_2$H$^+$ along the declination axis presents a curvy structure in the northern part, implying the acceleration of gas toward the central sub-clump, as well as the smaller-scale velocity gradient around cores, implying infall. 

    \item More direct signs of infall are found in the HNC line from the blue-asymmetry profiles. Among the 17 cores with the detections of HNC, nine show the infall signature. By using the Hill5 model, the infall velocity is estimated to be between 0.28 and 1.45 km\,s$^{-1}$. The mass infall rate of the most massive core ($M=10.2\,M_\odot$) is estimated as 2.9$\times$10$^{-3}\,M_\odot$\,yr$^{-1}$, while the remaining are $\sim$0.3--8$\times$10$^{-4}\,M_\odot$\,yr$^{-1}$. 

    \item The derived infall velocity has a correlation with the nonthermal velocity dispersion and the Mach number derived from optically thin lines. This implies that the nonthermal velocity dispersion does not purely trace the internal turbulence, but it is contaminated with infall motions. We also find a moderate correlation between the infall velocity with the core masses and the free-fall time of the cores. Considering these correlations and the fact that the timescale estimated from the velocity gradient of N$_2$H$^+$ is comparable to the free fall time, the cores are under gravitational collapse rather than supported by magnetic fields. 

    \item Two intermediate-mass cores ($M=10.2$ and 4.1 $\,M_\odot$) at the cluster center have higher infall velocities and mass infall rates than the other low-mass cores. We find a strong anti-correlation between the mass infall rate and the inverse mass-weighted distance, with higher-mass cores closer to the cluster center having higher mass infall rates, consistent with clump-fed scenarios. 

    \item A comparison of the infall velocity and mass infall rate in G337 with those in low-mass star-forming regions shows that both properties are significantly larger in G337, in spite of having similar core masses. This implies a more dynamical environment in G337, which can affect star formation. Indeed, such high-mass infall rate can contribute to make high-mass cores from intermediate-mass cores through gas feeding from the surrounding within a core free-fall time. 
\end{enumerate}

This study reports, in an infrared dark cloud, the dynamic infall both in the clump-scale and core-scale, the high infall velocity and infall rate from nine intermediate- and low-mass cores, and the higher infall rate around cores in high-mass star-forming regions than in low-mass star-forming regions. These findings propose that these infall motions play an important role in the formation of high-mass stars, supporting the core-growth scenario. With a larger sample of such studies, we would be able to obtain general conclusions beyond what is seen only in G337. 

\begin{acknowledgments} 
We thank an anonymous referee for constructive comments that helped improve this paper. 
K.M. is financially supported by Grants-in-Aid for the Japan Society for the Promotion of Science (JSPS) Fellows (KAKENHI Number JP22J21529), by FoPM, WINGS Program, the University of Tokyo, and an International Laboratory for astrophysics, neutrinos and cosmology Experiments (ILANCE). P.S. was partially supported by a Grant-in-Aid for Scientific Research (KAKENHI Number JP22H01271 and JP23H01221) of JSPS. P.S. was supported by Yoshinori Ohsumi Fund (Yoshinori Ohsumi Award for Fundamental Research).  
G.G. gratefully acknowledges support by the ANID BASAL project FB210003. 
Data analysis was in part carried out on the Multi-wavelength Data Analysis System operated by the Astronomy Data Center (ADC), National Astronomical Observatory of Japan. 
This paper uses the following ALMA data: ADS/JAO.ALMA\#2018.1.00299.S.  
ALMA is a partnership of ESO (representing its member states), NSF (USA) and NINS (Japan), together with NRC (Canada), $MOST$ and ASIAA (Taiwan), and KASI (Republic of Korea), in cooperation with the Republic of Chile. The Joint ALMA Observatory is operated by ESO, AUI/NRAO, and NAOJ. 
\facility {ALMA} 
\software{CASA \citep[][]{CASA22}, Numpy \citep{Harris20_numpy}, Scipy \citep{Virtanen_scipy}, Astropy \citep{Astropy_13, Astropy18, Astropy22}, Matplotlib \citep{Hunter_matplotlib}, bettermoments \citep{Teague18}, emcee \citep{Foreman-Mackey-emcee}, seaborn \citep{Waskom2021_seaborn}.}
\end{acknowledgments}

\bibliography{reference}

\begin{thebibliography}{}
\expandafter\ifx\csname natexlab\endcsname\relax\def\natexlab#1{#1}\fi
\providecommand{\url}[1]{\href{#1}{#1}}
\providecommand{\dodoi}[1]{doi:~\href{http://doi.org/#1}{\nolinkurl{#1}}}
\providecommand{\doeprint}[1]{\href{http://ascl.net/#1}{\nolinkurl{http://ascl.net/#1}}}
\providecommand{\doarXiv}[1]{\href{https://arxiv.org/abs/#1}{\nolinkurl{https://arxiv.org/abs/#1}}}

\bibitem[{{Ahmadi} {et~al.}(2023){Ahmadi}, {Beuther}, {Bosco}, {Gieser}, {Suri}, {Mottram}, {Kuiper}, {Henning}, {S{\'a}nchez-Monge}, {Linz}, {Pudritz}, {Semenov}, {Winters}, {M{\"o}ller}, {Beltr{\'a}n}, {Csengeri}, {Galv{\'a}n-Madrid}, {Johnston}, {Keto}, {Klaassen}, {Leurini}, {Longmore}, {Lumsden}, {Maud}, {Moscadelli}, {Palau}, {Peters}, {Ragan}, {Urquhart}, {Zhang}, \& {Zinnecker}}]{Ahmadi23}
{Ahmadi}, A., {Beuther}, H., {Bosco}, F., {et~al.} 2023, \aap, 677, A171, \dodoi{10.1051/0004-6361/202245580}

\bibitem[{{{\'A}lvarez-Guti{\'e}rrez} {et~al.}(2024){{\'A}lvarez-Guti{\'e}rrez}, {Stutz}, {Sandoval-Garrido}, {Louvet}, {Motte}, {Galv{\'a}n-Madrid}, {Cunningham}, {Sanhueza}, {Bonfand}, {Bontemps}, {Gusdorf}, {Ginsburg}, {Csengeri}, {Reyes}, {Salinas}, {Baug}, {Bronfman}, {Busquet}, {D{\'\i}az-Gonz{\'a}lez}, {Fernandez-Lopez}, {Guzm{\'a}n}, {Koley}, {Liu}, {Olguin}, {Valeille-Manet}, \& {Wyrowski}}]{almaimf_n2hp_24_arxiv}
{{\'A}lvarez-Guti{\'e}rrez}, R.~H., {Stutz}, A.~M., {Sandoval-Garrido}, N., {et~al.} 2024, arXiv e-prints, arXiv:2404.07363, \dodoi{10.48550/arXiv.2404.07363}

\bibitem[{{Astropy Collaboration} {et~al.}(2013){Astropy Collaboration}, {Robitaille}, {Tollerud}, {Greenfield}, {Droettboom}, {Bray}, {Aldcroft}, {Davis}, {Ginsburg}, {Price-Whelan}, {Kerzendorf}, {Conley}, {Crighton}, {Barbary}, {Muna}, {Ferguson}, {Grollier}, {Parikh}, {Nair}, {Unther}, {Deil}, {Woillez}, {Conseil}, {Kramer}, {Turner}, {Singer}, {Fox}, {Weaver}, {Zabalza}, {Edwards}, {Azalee Bostroem}, {Burke}, {Casey}, {Crawford}, {Dencheva}, {Ely}, {Jenness}, {Labrie}, {Lim}, {Pierfederici}, {Pontzen}, {Ptak}, {Refsdal}, {Servillat}, \& {Streicher}}]{Astropy_13}
{Astropy Collaboration}, {Robitaille}, T.~P., {Tollerud}, E.~J., {et~al.} 2013, \aap, 558, A33, \dodoi{10.1051/0004-6361/201322068}

\bibitem[{{Astropy Collaboration} {et~al.}(2018){Astropy Collaboration}, {Price-Whelan}, {Sip{\H{o}}cz}, {G{\"u}nther}, {Lim}, {Crawford}, {Conseil}, {Shupe}, {Craig}, {Dencheva}, {Ginsburg}, {VanderPlas}, {Bradley}, {P{\'e}rez-Su{\'a}rez}, {de Val-Borro}, {Aldcroft}, {Cruz}, {Robitaille}, {Tollerud}, {Ardelean}, {Babej}, {Bach}, {Bachetti}, {Bakanov}, {Bamford}, {Barentsen}, {Barmby}, {Baumbach}, {Berry}, {Biscani}, {Boquien}, {Bostroem}, {Bouma}, {Brammer}, {Bray}, {Breytenbach}, {Buddelmeijer}, {Burke}, {Calderone}, {Cano Rodr{\'\i}guez}, {Cara}, {Cardoso}, {Cheedella}, {Copin}, {Corrales}, {Crichton}, {D'Avella}, {Deil}, {Depagne}, {Dietrich}, {Donath}, {Droettboom}, {Earl}, {Erben}, {Fabbro}, {Ferreira}, {Finethy}, {Fox}, {Garrison}, {Gibbons}, {Goldstein}, {Gommers}, {Greco}, {Greenfield}, {Groener}, {Grollier}, {Hagen}, {Hirst}, {Homeier}, {Horton}, {Hosseinzadeh}, {Hu}, {Hunkeler}, {Ivezi{\'c}}, {Jain}, {Jenness}, {Kanarek}, {Kendrew}, {Kern}, {Kerzendorf}, {Khvalko}, {King}, {Kirkby}, {Kulkarni},
  {Kumar}, {Lee}, {Lenz}, {Littlefair}, {Ma}, {Macleod}, {Mastropietro}, {McCully}, {Montagnac}, {Morris}, {Mueller}, {Mumford}, {Muna}, {Murphy}, {Nelson}, {Nguyen}, {Ninan}, {N{\"o}the}, {Ogaz}, {Oh}, {Parejko}, {Parley}, {Pascual}, {Patil}, {Patil}, {Plunkett}, {Prochaska}, {Rastogi}, {Reddy Janga}, {Sabater}, {Sakurikar}, {Seifert}, {Sherbert}, {Sherwood-Taylor}, {Shih}, {Sick}, {Silbiger}, {Singanamalla}, {Singer}, {Sladen}, {Sooley}, {Sornarajah}, {Streicher}, {Teuben}, {Thomas}, {Tremblay}, {Turner}, {Terr{\'o}n}, {van Kerkwijk}, {de la Vega}, {Watkins}, {Weaver}, {Whitmore}, {Woillez}, {Zabalza}, \& {Astropy Contributors}}]{Astropy18}
{Astropy Collaboration}, {Price-Whelan}, A.~M., {Sip{\H{o}}cz}, B.~M., {et~al.} 2018, \aj, 156, 123, \dodoi{10.3847/1538-3881/aabc4f}

\bibitem[{{Astropy Collaboration} {et~al.}(2022){Astropy Collaboration}, {Price-Whelan}, {Lim}, {Earl}, {Starkman}, {Bradley}, {Shupe}, {Patil}, {Corrales}, {Brasseur}, {N{\"o}the}, {Donath}, {Tollerud}, {Morris}, {Ginsburg}, {Vaher}, {Weaver}, {Tocknell}, {Jamieson}, {van Kerkwijk}, {Robitaille}, {Merry}, {Bachetti}, {G{\"u}nther}, {Aldcroft}, {Alvarado-Montes}, {Archibald}, {B{\'o}di}, {Bapat}, {Barentsen}, {Baz{\'a}n}, {Biswas}, {Boquien}, {Burke}, {Cara}, {Cara}, {Conroy}, {Conseil}, {Craig}, {Cross}, {Cruz}, {D'Eugenio}, {Dencheva}, {Devillepoix}, {Dietrich}, {Eigenbrot}, {Erben}, {Ferreira}, {Foreman-Mackey}, {Fox}, {Freij}, {Garg}, {Geda}, {Glattly}, {Gondhalekar}, {Gordon}, {Grant}, {Greenfield}, {Groener}, {Guest}, {Gurovich}, {Handberg}, {Hart}, {Hatfield-Dodds}, {Homeier}, {Hosseinzadeh}, {Jenness}, {Jones}, {Joseph}, {Kalmbach}, {Karamehmetoglu}, {Ka{\l}uszy{\'n}ski}, {Kelley}, {Kern}, {Kerzendorf}, {Koch}, {Kulumani}, {Lee}, {Ly}, {Ma}, {MacBride}, {Maljaars}, {Muna}, {Murphy}, {Norman},
  {O'Steen}, {Oman}, {Pacifici}, {Pascual}, {Pascual-Granado}, {Patil}, {Perren}, {Pickering}, {Rastogi}, {Roulston}, {Ryan}, {Rykoff}, {Sabater}, {Sakurikar}, {Salgado}, {Sanghi}, {Saunders}, {Savchenko}, {Schwardt}, {Seifert-Eckert}, {Shih}, {Jain}, {Shukla}, {Sick}, {Simpson}, {Singanamalla}, {Singer}, {Singhal}, {Sinha}, {Sip{\H{o}}cz}, {Spitler}, {Stansby}, {Streicher}, {{\v{S}}umak}, {Swinbank}, {Taranu}, {Tewary}, {Tremblay}, {de Val-Borro}, {Van Kooten}, {Vasovi{\'c}}, {Verma}, {de Miranda Cardoso}, {Williams}, {Wilson}, {Winkel}, {Wood-Vasey}, {Xue}, {Yoachim}, {Zhang}, {Zonca}, \& {Astropy Project Contributors}}]{Astropy22}
{Astropy Collaboration}, {Price-Whelan}, A.~M., {Lim}, P.~L., {et~al.} 2022, \apj, 935, 167, \dodoi{10.3847/1538-4357/ac7c74}

\bibitem[{{Barnes} {et~al.}(2023){Barnes}, {Liu}, {Zhang}, {Tan}, {Bigiel}, {Caselli}, {Cosentino}, {Fontani}, {Henshaw}, {Jim{\'e}nez-Serra}, {Kalb}, {Law}, {Longmore}, {Parker}, {Pineda}, {S{\'a}nchez-Monge}, {Lim}, \& {Wang}}]{Barnes23}
{Barnes}, A.~T., {Liu}, J., {Zhang}, Q., {et~al.} 2023, \aap, 675, A53, \dodoi{10.1051/0004-6361/202245668}

\bibitem[{{Bjerkeli} {et~al.}(2016){Bjerkeli}, {J{\o}rgensen}, \& {Brinch}}]{Bjerkeli16}
{Bjerkeli}, P., {J{\o}rgensen}, J.~K., \& {Brinch}, C. 2016, \aap, 587, A145, \dodoi{10.1051/0004-6361/201527310}

\bibitem[{{Campbell} {et~al.}(2016){Campbell}, {Friesen}, {Martin}, {Caselli}, {Kauffmann}, \& {Pineda}}]{Campbell16}
{Campbell}, J.~L., {Friesen}, R.~K., {Martin}, P.~G., {et~al.} 2016, \apj, 819, 143, \dodoi{10.3847/0004-637X/819/2/143}

\bibitem[{{CASA Team} {et~al.}(2022){CASA Team}, {Bean}, {Bhatnagar}, {Castro}, {Donovan Meyer}, {Emonts}, {Garcia}, {Garwood}, {Golap}, {Gonzalez Villalba}, {Harris}, {Hayashi}, {Hoskins}, {Hsieh}, {Jagannathan}, {Kawasaki}, {Keimpema}, {Kettenis}, {Lopez}, {Marvil}, {Masters}, {McNichols}, {Mehringer}, {Miel}, {Moellenbrock}, {Montesino}, {Nakazato}, {Ott}, {Petry}, {Pokorny}, {Raba}, {Rau}, {Schiebel}, {Schweighart}, {Sekhar}, {Shimada}, {Small}, {Steeb}, {Sugimoto}, {Suoranta}, {Tsutsumi}, {van Bemmel}, {Verkouter}, {Wells}, {Xiong}, {Szomoru}, {Griffith}, {Glendenning}, \& {Kern}}]{CASA22}
{CASA Team}, {Bean}, B., {Bhatnagar}, S., {et~al.} 2022, \pasp, 134, 114501, \dodoi{10.1088/1538-3873/ac9642}

\bibitem[{{Chen} {et~al.}(2019){Chen}, {Zhang}, {Wright}, {Busquet}, {Lin}, {Liu}, {Olguin}, {Sanhueza}, {Nakamura}, {Palau}, {Ohashi}, {Tatematsu}, \& {Liao}}]{Chen19}
{Chen}, H.-R.~V., {Zhang}, Q., {Wright}, M.~C.~H., {et~al.} 2019, \apj, 875, 24, \dodoi{10.3847/1538-4357/ab0f3e}

\bibitem[{{Chira} {et~al.}(2014){Chira}, {Smith}, {Klessen}, {Stutz}, \& {Shetty}}]{Chira14}
{Chira}, R.-A., {Smith}, R.~J., {Klessen}, R.~S., {Stutz}, A.~M., \& {Shetty}, R. 2014, \mnras, 444, 874, \dodoi{10.1093/mnras/stu1497}

\bibitem[{{Contreras}(2018)}]{Contreas_yclean_18}
{Contreras}, Y. 2018, {Automatic Line Clean}, 1.0,  Zenodo, \dodoi{10.5281/zenodo.1216881}

\bibitem[{{Contreras} {et~al.}(2018){Contreras}, {Sanhueza}, {Jackson}, {Guzm{\'a}n}, {Longmore}, {Garay}, {Zhang}, {Nguyễn-Lu'o'ng}, {Tatematsu}, {Nakamura}, {Sakai}, {Ohashi}, {Liu}, {Saito}, {Gomez}, {Rathborne}, \& {Whitaker}}]{Contreras18}
{Contreras}, Y., {Sanhueza}, P., {Jackson}, J.~M., {et~al.} 2018, \apj, 861, 14, \dodoi{10.3847/1538-4357/aac2ec}

\bibitem[{{Csengeri} {et~al.}(2011){Csengeri}, {Bontemps}, {Schneider}, {Motte}, \& {Dib}}]{Csengeri11}
{Csengeri}, T., {Bontemps}, S., {Schneider}, N., {Motte}, F., \& {Dib}, S. 2011, \aap, 527, A135, \dodoi{10.1051/0004-6361/201014984}

\bibitem[{{De Vries} \& {Myers}(2005)}]{De_Vries05}
{De Vries}, C.~H., \& {Myers}, P.~C. 2005, \apj, 620, 800, \dodoi{10.1086/427141}

\bibitem[{{Evans}(1999)}]{Evans99}
{Evans}, Neal~J., I. 1999, \araa, 37, 311, \dodoi{10.1146/annurev.astro.37.1.311}

\bibitem[{{Foreman-Mackey} {et~al.}(2013){Foreman-Mackey}, {Hogg}, {Lang}, \& {Goodman}}]{Foreman-Mackey-emcee}
{Foreman-Mackey}, D., {Hogg}, D.~W., {Lang}, D., \& {Goodman}, J. 2013, \pasp, 125, 306, \dodoi{10.1086/670067}

\bibitem[{{Fuller} {et~al.}(2005){Fuller}, {Williams}, \& {Sridharan}}]{Fuller05}
{Fuller}, G.~A., {Williams}, S.~J., \& {Sridharan}, T.~K. 2005, \aap, 442, 949, \dodoi{10.1051/0004-6361:20042110}

\bibitem[{{Goodman} {et~al.}(1993){Goodman}, {Benson}, {Fuller}, \& {Myers}}]{Goodman93}
{Goodman}, A.~A., {Benson}, P.~J., {Fuller}, G.~A., \& {Myers}, P.~C. 1993, \apj, 406, 528, \dodoi{10.1086/172465}

\bibitem[{{Guzm{\'a}n} {et~al.}(2015){Guzm{\'a}n}, {Sanhueza}, {Contreras}, {Smith}, {Jackson}, {Hoq}, \& {Rathborne}}]{Guzman15}
{Guzm{\'a}n}, A.~E., {Sanhueza}, P., {Contreras}, Y., {et~al.} 2015, \apj, 815, 130, \dodoi{10.1088/0004-637X/815/2/130}

\bibitem[{Harris {et~al.}(2020)Harris, Millman, van~der Walt, Gommers, Virtanen, Cournapeau, Wieser, Taylor, Berg, Smith, Kern, Picus, Hoyer, van Kerkwijk, Brett, Haldane, del Río, Wiebe, Peterson, Gérard-Marchant, Sheppard, Reddy, Weckesser, Abbasi, Gohlke, \& Oliphant}]{Harris20_numpy}
Harris, C.~R., Millman, K.~J., van~der Walt, S.~J., {et~al.} 2020, Nature, 585, 357, \dodoi{10.1038/s41586-020-2649-2}

\bibitem[{{Henshaw} {et~al.}(2014){Henshaw}, {Caselli}, {Fontani}, {Jim{\'e}nez-Serra}, \& {Tan}}]{Henshaw14}
{Henshaw}, J.~D., {Caselli}, P., {Fontani}, F., {Jim{\'e}nez-Serra}, I., \& {Tan}, J.~C. 2014, \mnras, 440, 2860, \dodoi{10.1093/mnras/stu446}

\bibitem[{{Hunter}(2007)}]{Hunter_matplotlib}
{Hunter}, J.~D. 2007, Computing in Science and Engineering, 9, 90, \dodoi{10.1109/MCSE.2007.55}

\bibitem[{{Izumi} {et~al.}(2024){Izumi}, {Sanhueza}, {Koch}, {Lu}, {Li}, {Sabatini}, {Olguin}, {Zhang}, {Nakamura}, {Tatematsu}, {Morii}, {Sakai}, \& {Tafoya}}]{Izumi24}
{Izumi}, N., {Sanhueza}, P., {Koch}, P.~M., {et~al.} 2024, \apj, 963, 163, \dodoi{10.3847/1538-4357/ad18c6}

\bibitem[{{Jackson} {et~al.}(2019){Jackson}, {Whitaker}, {Rathborne}, {Foster}, {Contreras}, {Sanhueza}, {Stephens}, {Longmore}, \& {Allingham}}]{Jackson19}
{Jackson}, J.~M., {Whitaker}, J.~S., {Rathborne}, J.~M., {et~al.} 2019, \apj, 870, 5, \dodoi{10.3847/1538-4357/aaef84}

\bibitem[{{Kim} {et~al.}(2021){Kim}, {Lee}, {Maheswar}, {Myers}, \& {Kim}}]{Kim21}
{Kim}, M.-R., {Lee}, C.~W., {Maheswar}, G., {Myers}, P.~C., \& {Kim}, G. 2021, \apj, 910, 112, \dodoi{10.3847/1538-4357/abe4d3}

\bibitem[{{Kong} {et~al.}(2017){Kong}, {Tan}, {Caselli}, {Fontani}, {Liu}, \& {Butler}}]{Kong17}
{Kong}, S., {Tan}, J.~C., {Caselli}, P., {et~al.} 2017, \apj, 834, 193, \dodoi{10.3847/1538-4357/834/2/193}

\bibitem[{{Lee} {et~al.}(2001){Lee}, {Myers}, \& {Tafalla}}]{Lee01}
{Lee}, C.~W., {Myers}, P.~C., \& {Tafalla}, M. 2001, \apjs, 136, 703, \dodoi{10.1086/322534}

\bibitem[{{Li} {et~al.}(2021){Li}, {Lu}, {Zhang}, {Lee}, {Sanhueza}, {Beuther}, \& {Jim{\'e}nez-Serra}}]{Li21a}
{Li}, S., {Lu}, X., {Zhang}, Q., {et~al.} 2021, \apjl, 912, L7, \dodoi{10.3847/2041-8213/abf64f}

\bibitem[{{Li} {et~al.}(2020){Li}, {Sanhueza}, {Zhang}, {Nakamura}, {Lu}, {Wang}, {Liu}, {Tatematsu}, {Jackson}, {Silva}, {Guzm{\'a}n}, {Sakai}, {Izumi}, {Tafoya}, {Li}, {Contreras}, {Morii}, \& {Kim}}]{Li20}
{Li}, S., {Sanhueza}, P., {Zhang}, Q., {et~al.} 2020, \apj, 903, 119, \dodoi{10.3847/1538-4357/abb81f}

\bibitem[{{Li} {et~al.}(2022){Li}, {Sanhueza}, {Lu}, {Lee}, {Zhang}, {Bovino}, {Sabatini}, {Liu}, {Kim}, {Morii}, {Tafoya}, {Tatematsu}, {Sakai}, {Wang}, {Li}, {Silva}, {Izumi}, \& {Allingham}}]{Li22}
{Li}, S., {Sanhueza}, P., {Lu}, X., {et~al.} 2022, \apj, 939, 102, \dodoi{10.3847/1538-4357/ac94d4}

\bibitem[{{Li} {et~al.}(2023){Li}, {Sanhueza}, {Zhang}, {Guido}, {Sabatini}, {Morii}, {Lu}, {Tafoya}, {Nakamura}, {Izumi}, {Tatematsu}, \& {Li}}]{Li23}
{Li}, S., {Sanhueza}, P., {Zhang}, Q., {et~al.} 2023, \apj, 949, 109, \dodoi{10.3847/1538-4357/acc58f}

\bibitem[{{Louvet} {et~al.}(2019){Louvet}, {Neupane}, {Garay}, {Russeil}, {Zavagno}, {Guzman}, {Gomez}, {Bronfman}, \& {Nony}}]{Louvet19}
{Louvet}, F., {Neupane}, S., {Garay}, G., {et~al.} 2019, \aap, 622, A99, \dodoi{10.1051/0004-6361/201732282}

\bibitem[{{Lu} {et~al.}(2015){Lu}, {Zhang}, {Wang}, \& {Gu}}]{Lu15}
{Lu}, X., {Zhang}, Q., {Wang}, K., \& {Gu}, Q. 2015, \apj, 805, 171, \dodoi{10.1088/0004-637X/805/2/171}

\bibitem[{{Lu} {et~al.}(2018){Lu}, {Zhang}, {Liu}, {Sanhueza}, {Tatematsu}, {Feng}, {Smith}, {Myers}, {Sridharan}, \& {Gu}}]{Lu18}
{Lu}, X., {Zhang}, Q., {Liu}, H.~B., {et~al.} 2018, \apj, 855, 9, \dodoi{10.3847/1538-4357/aaad11}

\bibitem[{{Mai} {et~al.}(2024){Mai}, {Liu}, {Liu}, {Zhu}, {Garay}, {Goldsmith}, {Juvela}, {Liu}, {Mannfors}, {Tej}, {Sanhueza}, {Li}, {Xu}, {Semadeni}, {Jiao}, {Peng}, {Baug}, {Yang}, {Dewangan}, {Bronfman}, {G{\'o}mez}, {Palau}, {Lee}, {Qin}, {Tatematsu}, {Chibueze}, {Yang}, {Lu}, {Luo}, {Gu}, {Issac}, {Zhang}, {Li}, {Zhang}, \& {T{\'o}th}}]{Mai24}
{Mai}, X., {Liu}, T., {Liu}, X., {et~al.} 2024, \apjl, 961, L35, \dodoi{10.3847/2041-8213/ad19c3}

\bibitem[{{Molet} {et~al.}(2019){Molet}, {Brouillet}, {Nony}, {Gusdorf}, {Motte}, {Despois}, {Louvet}, {Bontemps}, \& {Herpin}}]{Molet19}
{Molet}, J., {Brouillet}, N., {Nony}, T., {et~al.} 2019, \aap, 626, A132, \dodoi{10.1051/0004-6361/201935497}

\bibitem[{{Morii} {et~al.}(2021){Morii}, {Sanhueza}, {Nakamura}, {Jackson}, {Li}, {Beuther}, {Zhang}, {Feng}, {Tafoya}, {Guzm{\'a}n}, {Izumi}, {Sakai}, {Lu}, {Tatematsu}, {Ohashi}, {Silva}, {Olguin}, \& {Contreras}}]{Morii21}
{Morii}, K., {Sanhueza}, P., {Nakamura}, F., {et~al.} 2021, \apj, 923, 147, \dodoi{10.3847/1538-4357/ac2365}

\bibitem[{{Morii} {et~al.}(2023){Morii}, {Sanhueza}, {Nakamura}, {Zhang}, {Sabatini}, {Beuther}, {Lu}, {Li}, {Garay}, {Jackson}, {Olguin}, {Tafoya}, {Tatematsu}, {Izumi}, {Sakai}, \& {Silva}}]{Morii23}
---. 2023, \apj, 950, 148, \dodoi{10.3847/1538-4357/acccea}

\bibitem[{{Morii} {et~al.}(2024){Morii}, {Sanhueza}, {Zhang}, {Nakamura}, {Li}, {Sabatini}, {Olguin}, {Beuther}, {Tafoya}, {Izumi}, {Tatematsu}, \& {Sakai}}]{Morii24}
{Morii}, K., {Sanhueza}, P., {Zhang}, Q., {et~al.} 2024, \apj, 966, 171, \dodoi{10.3847/1538-4357/ad32d0}

\bibitem[{{Nony} {et~al.}(2018){Nony}, {Louvet}, {Motte}, {Molet}, {Marsh}, {Chapillon}, {Gusdorf}, {Brouillet}, {Bontemps}, {Csengeri}, {Despois}, {Nguyen Luong}, {Duarte-Cabral}, \& {Maury}}]{Nony18}
{Nony}, T., {Louvet}, F., {Motte}, F., {et~al.} 2018, \aap, 618, L5, \dodoi{10.1051/0004-6361/201833863}

\bibitem[{{Ohashi} {et~al.}(2016){Ohashi}, {Sanhueza}, {Chen}, {Zhang}, {Busquet}, {Nakamura}, {Palau}, \& {Tatematsu}}]{Ohashi16}
{Ohashi}, S., {Sanhueza}, P., {Chen}, H.-R.~V., {et~al.} 2016, \apj, 833, 209, \dodoi{10.3847/1538-4357/833/2/209}

\bibitem[{{Olguin} {et~al.}(2023){Olguin}, {Sanhueza}, {Chen}, {Lu}, {Oya}, {Zhang}, {Ginsburg}, {Taniguchi}, {Li}, {Morii}, {Sakai}, \& {Nakamura}}]{Olguin23}
{Olguin}, F.~A., {Sanhueza}, P., {Chen}, H.-R.~V., {et~al.} 2023, \apjl, 959, L31, \dodoi{10.3847/2041-8213/ad1100}

\bibitem[{{Peretto} {et~al.}(2014){Peretto}, {Fuller}, {Andr{\'e}}, {Arzoumanian}, {Rivilla}, {Bardeau}, {Duarte Puertas}, {Guzman Fernandez}, {Lenfestey}, {Li}, {Olguin}, {R{\"o}ck}, {de Villiers}, \& {Williams}}]{Peretto14}
{Peretto}, N., {Fuller}, G.~A., {Andr{\'e}}, P., {et~al.} 2014, \aap, 561, A83, \dodoi{10.1051/0004-6361/201322172}

\bibitem[{{Peretto, N.} {et~al.}(2006){Peretto, N.}, {Andr{\'e}, Ph.}, \& {Belloche, A.}}]{Peretto06}
{Peretto, N.}, {Andr{\'e}, Ph.}, \& {Belloche, A.} 2006, A\&A, 445, 979, \dodoi{10.1051/0004-6361:20053324}

\bibitem[{{Pillai} {et~al.}(2019){Pillai}, {Kauffmann}, {Zhang}, {Sanhueza}, {Leurini}, {Wang}, {Sridharan}, \& {K{\"o}nig}}]{Pillai19}
{Pillai}, T., {Kauffmann}, J., {Zhang}, Q., {et~al.} 2019, \aap, 622, A54, \dodoi{10.1051/0004-6361/201732570}

\bibitem[{{Pouteau} {et~al.}(2023){Pouteau}, {Motte}, {Nony}, {Gonz{\'a}lez}, {Joncour}, {Robitaille}, {Busquet}, {Galv{\'a}n-Madrid}, {Gusdorf}, {Hennebelle}, {Ginsburg}, {Csengeri}, {Sanhueza}, {Dell'Ova}, {Stutz}, {Towner}, {Cunningham}, {Louvet}, {Men'shchikov}, {Fern{\'a}ndez-L{\'o}pez}, {Schneider}, {Armante}, {Bally}, {Baug}, {Bonfand}, {Bontemps}, {Bronfman}, {Brouillet}, {D{\'\i}az-Gonz{\'a}lez}, {Herpin}, {Lefloch}, {Liu}, {Lu}, {Nakamura}, {Nguyen-Luong}, {Olguin}, {Tatematsu}, \& {Valeille-Manet}}]{Pouteau23}
{Pouteau}, Y., {Motte}, F., {Nony}, T., {et~al.} 2023, \aap, 674, A76, \dodoi{10.1051/0004-6361/202244776}

\bibitem[{{Rathborne} {et~al.}(2016){Rathborne}, {Whitaker}, {Jackson}, {Foster}, {Contreras}, {Stephens}, {Guzm{\'a}n}, {Longmore}, {Sanhueza}, {Schuller}, {Wyrowski}, \& {Urquhart}}]{Rathborne16}
{Rathborne}, J.~M., {Whitaker}, J.~S., {Jackson}, J.~M., {et~al.} 2016, \pasa, 33, e030, \dodoi{10.1017/pasa.2016.23}

\bibitem[{{Redaelli} {et~al.}(2022){Redaelli}, {Bovino}, {Sanhueza}, {Morii}, {Sabatini}, {Caselli}, {Giannetti}, \& {Li}}]{Redaelli22}
{Redaelli}, E., {Bovino}, S., {Sanhueza}, P., {et~al.} 2022, \apj, 936, 169, \dodoi{10.3847/1538-4357/ac85b4}

\bibitem[{{Reiter} {et~al.}(2011){Reiter}, {Shirley}, {Wu}, {Brogan}, {Wootten}, \& {Tatematsu}}]{Reiter11}
{Reiter}, M., {Shirley}, Y.~L., {Wu}, J., {et~al.} 2011, \apj, 740, 40, \dodoi{10.1088/0004-637X/740/1/40}

\bibitem[{{Rigby} {et~al.}(2024){Rigby}, {Peretto}, {Anderson}, {Ragan}, {Priestley}, {Fuller}, {Thompson}, {Traficante}, {Watkins}, \& {Williams}}]{Rigby24}
{Rigby}, A.~J., {Peretto}, N., {Anderson}, M., {et~al.} 2024, \mnras, 528, 1172, \dodoi{10.1093/mnras/stae030}

\bibitem[{{Rygl} {et~al.}(2013){Rygl}, {Wyrowski}, {Schuller}, \& {Menten}}]{Rygl13}
{Rygl}, K.~L.~J., {Wyrowski}, F., {Schuller}, F., \& {Menten}, K.~M. 2013, \aap, 549, A5, \dodoi{10.1051/0004-6361/201219574}

\bibitem[{{Sanhueza} {et~al.}(2010){Sanhueza}, {Garay}, {Bronfman}, {Mardones}, {May}, \& {Saito}}]{Sanhueza10}
{Sanhueza}, P., {Garay}, G., {Bronfman}, L., {et~al.} 2010, \apj, 715, 18, \dodoi{10.1088/0004-637X/715/1/18}

\bibitem[{{Sanhueza} {et~al.}(2012){Sanhueza}, {Jackson}, {Foster}, {Garay}, {Silva}, \& {Finn}}]{Sanhueza12}
{Sanhueza}, P., {Jackson}, J.~M., {Foster}, J.~B., {et~al.} 2012, \apj, 756, 60, \dodoi{10.1088/0004-637X/756/1/60}

\bibitem[{{Sanhueza} {et~al.}(2013){Sanhueza}, {Jackson}, {Foster}, {Jimenez-Serra}, {Dirienzo}, \& {Pillai}}]{Sanhueza13}
---. 2013, \apj, 773, 123, \dodoi{10.1088/0004-637X/773/2/123}

\bibitem[{{Sanhueza} {et~al.}(2017){Sanhueza}, {Jackson}, {Zhang}, {Guzm{\'a}n}, {Lu}, {Stephens}, {Wang}, \& {Tatematsu}}]{Sanhueza17}
{Sanhueza}, P., {Jackson}, J.~M., {Zhang}, Q., {et~al.} 2017, \apj, 841, 97, \dodoi{10.3847/1538-4357/aa6ff8}

\bibitem[{{Sanhueza} {et~al.}(2019){Sanhueza}, {Contreras}, {Wu}, {Jackson}, {Guzm{\'a}n}, {Zhang}, {Li}, {Lu}, {Silva}, {Izumi}, {Liu}, {Miura}, {Tatematsu}, {Sakai}, {Beuther}, {Garay}, {Ohashi}, {Saito}, {Nakamura}, {Saigo}, {Veena}, {Nguyen-Luong}, \& {Tafoya}}]{Sanhueza19}
{Sanhueza}, P., {Contreras}, Y., {Wu}, B., {et~al.} 2019, \apj, 886, 102, \dodoi{10.3847/1538-4357/ab45e9}

\bibitem[{{Sanhueza} {et~al.}(2021){Sanhueza}, {Girart}, {Padovani}, {Galli}, {Hull}, {Zhang}, {Cortes}, {Stephens}, {Fern{\'a}ndez-L{\'o}pez}, {Jackson}, {Frau}, {Kock}, {Wu}, {Zapata}, {Olguin}, {Lu}, {Silva}, {Tang}, {Sakai}, {Guzm{\'a}n}, {Tatematsu}, {Nakamura}, \& {Chen}}]{Sanhueza21}
{Sanhueza}, P., {Girart}, J.~M., {Padovani}, M., {et~al.} 2021, \apjl, 915, L10, \dodoi{10.3847/2041-8213/ac081c}

\bibitem[{{Schneider} {et~al.}(2010){Schneider}, {Csengeri}, {Bontemps}, {Motte}, {Simon}, {Hennebelle}, {Federrath}, \& {Klessen}}]{Schneider10}
{Schneider}, N., {Csengeri}, T., {Bontemps}, S., {et~al.} 2010, \aap, 520, A49, \dodoi{10.1051/0004-6361/201014481}

\bibitem[{{Smith} {et~al.}(2012){Smith}, {Shetty}, {Stutz}, \& {Klessen}}]{Smith12}
{Smith}, R.~J., {Shetty}, R., {Stutz}, A.~M., \& {Klessen}, R.~S. 2012, \apj, 750, 64, \dodoi{10.1088/0004-637X/750/1/64}

\bibitem[{{Svoboda} {et~al.}(2019){Svoboda}, {Shirley}, {Traficante}, {Battersby}, {Fuller}, {Zhang}, {Beuther}, {Peretto}, {Brogan}, \& {Hunter}}]{Svoboda19}
{Svoboda}, B.~E., {Shirley}, Y.~L., {Traficante}, A., {et~al.} 2019, \apj, 886, 36, \dodoi{10.3847/1538-4357/ab40ca}

\bibitem[{{Teague} \& {Foreman-Mackey}(2018)}]{Teague18}
{Teague}, R., \& {Foreman-Mackey}, D. 2018, Research Notes of the American Astronomical Society, 2, 173, \dodoi{10.3847/2515-5172/aae265}

\bibitem[{{Tobin} {et~al.}(2011){Tobin}, {Hartmann}, {Chiang}, {Looney}, {Bergin}, {Chandler}, {Masqu{\'e}}, {Maret}, \& {Heitsch}}]{Tobin11}
{Tobin}, J.~J., {Hartmann}, L., {Chiang}, H.-F., {et~al.} 2011, \apj, 740, 45, \dodoi{10.1088/0004-637X/740/1/45}

\bibitem[{{Traficante} {et~al.}(2017){Traficante}, {Fuller}, {Billot}, {Duarte-Cabral}, {Merello}, {Molinari}, {Peretto}, \& {Schisano}}]{Traficante17}
{Traficante}, A., {Fuller}, G.~A., {Billot}, N., {et~al.} 2017, \mnras, 470, 3882, \dodoi{10.1093/mnras/stx1375}

\bibitem[{{Traficante} {et~al.}(2018){Traficante}, {Fuller}, {Smith}, {Billot}, {Duarte-Cabral}, {Peretto}, {Molinari}, \& {Pineda}}]{Traficante18}
{Traficante}, A., {Fuller}, G.~A., {Smith}, R.~J., {et~al.} 2018, \mnras, 473, 4975, \dodoi{10.1093/mnras/stx2672}

\bibitem[{{Traficante} {et~al.}(2023){Traficante}, {Jones}, {Avison}, {Fuller}, {Benedettini}, {Elia}, {Molinari}, {Peretto}, {Pezzuto}, {Pillai}, {Rygl}, {Schisano}, \& {Smith}}]{Traficante23}
{Traficante}, A., {Jones}, B.~M., {Avison}, A., {et~al.} 2023, \mnras, 520, 2306, \dodoi{10.1093/mnras/stad272}

\bibitem[{Virtanen {et~al.}(2020)Virtanen, Gommers, Oliphant, Haberland, Reddy, Cournapeau, Burovski, Peterson, Weckesser, Bright, van~der Walt, Brett, Wilson, Millman, Mayorov, Nelson, Jones, Kern, Larson, Carey, Polat, Feng, Moore, VanderPlas, Laxalde, Perktold, Cimrman, Henriksen, Quintero, Harris, Archibald, Ribeiro, Pedregosa, van Mulbregt, Vijaykumar, Bardelli, Rothberg, Hilboll, Kloeckner, Scopatz, Lee, Rokem, Woods, Fulton, Masson, Häggström, Fitzgerald, Nicholson, Hagen, Pasechnik, Olivetti, Martin, Wieser, Silva, Lenders, Wilhelm, Young, Price, Ingold, Allen, Lee, Audren, Probst, Dietrich, Silterra, Webber, Slavič, Nothman, Buchner, Kulick, Schönberger, de~Miranda~Cardoso, Reimer, Harrington, Rodríguez, Nunez-Iglesias, Kuczynski, Tritz, Thoma, Newville, Kümmerer, Bolingbroke, Tartre, Pak, Smith, Nowaczyk, Shebanov, Pavlyk, Brodtkorb, Lee, McGibbon, Feldbauer, Lewis, Tygier, Sievert, Vigna, Peterson, More, Pudlik, Oshima, Pingel, Robitaille, Spura, Jones, Cera, Leslie, Zito, Krauss, Upadhyay,
  Halchenko, Vázquez-Baeza, \& {SciPy 1.0 Contributors}}]{Virtanen_scipy}
Virtanen, P., Gommers, R., Oliphant, T.~E., {et~al.} 2020, Nature Methods, 17, 261, \dodoi{10.1038/s41592-019-0686-2}

\bibitem[{Waskom(2021)}]{Waskom2021_seaborn}
Waskom, M.~L. 2021, Journal of Open Source Software, 6, 3021, \dodoi{10.21105/joss.03021}

\bibitem[{{Wells} {et~al.}(2024){Wells}, {Beuther}, {Molinari}, {Schilke}, {Battersby}, {Ho}, {S{\'a}nchez-Monge}, {Jones}, {Scheuck}, {Syed}, {Gieser}, {Kuiper}, {Elia}, {Coletta}, {Traficante}, {Wallace}, {Rigby}, {Klessen}, {Zhang}, {Walch}, {Beltr{\'a}n}, {Tang}, {Fuller}, {Lis}, {M{\"o}ller}, {van der Tak}, {Klaassen}, {Clarke}, {Moscadelli}, {Mininni}, {Zinnecker}, {Maruccia}, {Pezzuto}, {Benedettini}, {Soler}, {Brogan}, {Avison}, {Sanhueza}, {Schisano}, {Liu}, {Fontani}, {Rygl}, {Wyrowski}, {Bally}, {Walker}, {Ahmadi}, {Koch}, {Merello}, {Law}, \& {Testi}}]{Wells24}
{Wells}, M.~R.~A., {Beuther}, H., {Molinari}, S., {et~al.} 2024, arXiv e-prints, arXiv:2408.08299, \dodoi{10.48550/arXiv.2408.08299}

\bibitem[{{Whitaker} {et~al.}(2017){Whitaker}, {Jackson}, {Rathborne}, {Foster}, {Contreras}, {Sanhueza}, {Stephens}, \& {Longmore}}]{Whitaker17}
{Whitaker}, J.~S., {Jackson}, J.~M., {Rathborne}, J.~M., {et~al.} 2017, \aj, 154, 140, \dodoi{10.3847/1538-3881/aa86ad}

\bibitem[{{Wu} {et~al.}(2010){Wu}, {Evans}, {Shirley}, \& {Knez}}]{WuEvans10}
{Wu}, J., {Evans}, Neal~J., I., {Shirley}, Y.~L., \& {Knez}, C. 2010, \apjs, 188, 313, \dodoi{10.1088/0067-0049/188/2/313}

\bibitem[{{Wyrowski} {et~al.}(2016){Wyrowski}, {G{\"u}sten}, {Menten}, {Wiesemeyer}, {Csengeri}, {Heyminck}, {Klein}, {K{\"o}nig}, \& {Urquhart}}]{Wyrowski16}
{Wyrowski}, F., {G{\"u}sten}, R., {Menten}, K.~M., {et~al.} 2016, \aap, 585, A149, \dodoi{10.1051/0004-6361/201526361}

\bibitem[{{Xie} {et~al.}(2021){Xie}, {Wu}, {Fuller}, {Peretto}, {Ren}, {Chen}, {Yan}, {Li}, {Duan}, {Xia}, {Wang}, \& {Li}}]{Jin-Jin21}
{Xie}, J.-J., {Wu}, J.-W., {Fuller}, G.~A., {et~al.} 2021, Research in Astronomy and Astrophysics, 21, 208, \dodoi{10.1088/1674-4527/21/8/208}

\bibitem[{{Xu} {et~al.}(2023{\natexlab{a}}){Xu}, {Wang}, {He}, {Wu}, {Zhu}, \& {Mardones}}]{Xu23_hcn}
{Xu}, F., {Wang}, K., {He}, Y., {et~al.} 2023{\natexlab{a}}, \apjs, 269, 38, \dodoi{10.3847/1538-4365/acfee2}

\bibitem[{{Xu} {et~al.}(2024){Xu}, {Wang}, {Liu}, {Zhu}, {Garay}, {Liu}, {Goldsmith}, {Zhang}, {Sanhueza}, {Qin}, {He}, {Juvela}, {Tej}, {Liu}, {Li}, {Morii}, {Zhang}, {Zhou}, {Stutz}, {Evans}, {Kim}, {Liu}, {Mardones}, {Li}, {Bronfman}, {Tatematsu}, {Lee}, {Lu}, {Mai}, {Jiao}, {Chibueze}, {Su}, \& {T{\'o}th}}]{Xu24}
{Xu}, F., {Wang}, K., {Liu}, T., {et~al.} 2024, Research in Astronomy and Astrophysics, 24, 065011, \dodoi{10.1088/1674-4527/ad3dc3}

\bibitem[{{Xu} {et~al.}(2023{\natexlab{b}}){Xu}, {Wang}, {Liu}, {Goldsmith}, {Zhang}, {Juvela}, {Liu}, {Qin}, {Li}, {Tej}, {Garay}, {Bronfman}, {Li}, {Wu}, {G{\'o}mez}, {V{\'a}zquez-Semadeni}, {Tatematsu}, {Ren}, {Zhang}, {Toth}, {Liu}, {Yue}, {Zhang}, {Baug}, {Issac}, {Stutz}, {Liu}, {Fuller}, {Tang}, {Zhang}, {Dewangan}, {Lee}, {Zhou}, {Xie}, {Jiao}, {Wang}, {Liu}, {Luo}, {Soam}, \& {Eswaraiah}}]{Xu23}
{Xu}, F.-W., {Wang}, K., {Liu}, T., {et~al.} 2023{\natexlab{b}}, \mnras, 520, 3259, \dodoi{10.1093/mnras/stad012}

\bibitem[{{Yang} {et~al.}(2021){Yang}, {Jiang}, {Chen}, {Ao}, \& {Yu}}]{Yang21}
{Yang}, Y., {Jiang}, Z., {Chen}, Z., {Ao}, Y., \& {Yu}, S. 2021, \apj, 922, 144, \dodoi{10.3847/1538-4357/ac22ab}

\bibitem[{{Yu} {et~al.}(2022){Yu}, {Jiang}, {Yang}, {Chen}, \& {Feng}}]{Yu22}
{Yu}, S., {Jiang}, Z., {Yang}, Y., {Chen}, Z., \& {Feng}, H. 2022, Research in Astronomy and Astrophysics, 22, 095014, \dodoi{10.1088/1674-4527/ac7d9d}

\bibitem[{{Zhang} \& {Wang}(2011)}]{Zhang11}
{Zhang}, Q., \& {Wang}, K. 2011, \apj, 733, 26, \dodoi{10.1088/0004-637X/733/1/26}

\end{thebibliography}
\bibliographystyle{aasjournal}

\appendix
\section{Extraction of outflow component}
\label{sec:Appendix_outflowextraction}
\begin{figure*}
    \centering
    \gridline{\fig{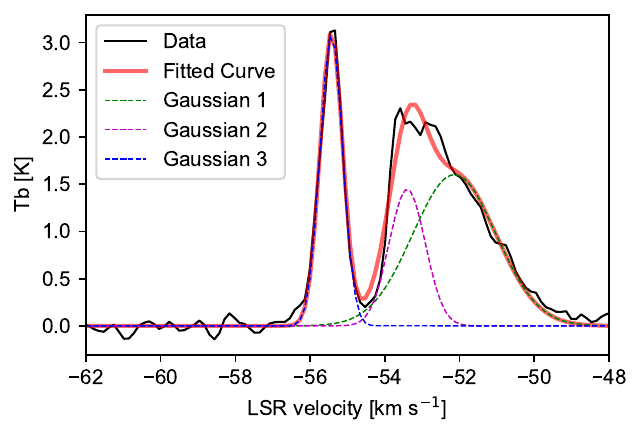}{0.47\textwidth}{(a) ALMA2} 
              \fig{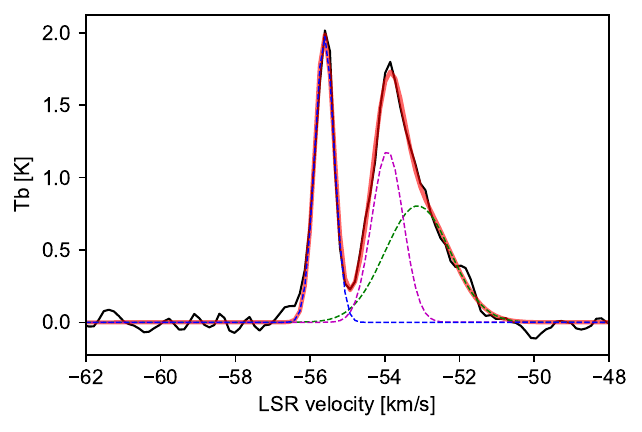}{0.47\textwidth}{(b) ALMA18}
             }
    \caption{Multiple gaussian fitting. }
    \label{fig:gaussfit}
\end{figure*}

We extracted the outflow component from the fitting of the line profile with a multiple Gaussian. 
The black lines in Figure~\ref{fig:gaussfit} are the line spectra of HNC averaged within each core, ALMA2 and ALMA18. 
In the red-shifted component, there are high-velocity tails, likely produced by high-velocity outflowing gas. 
We fit the spectra with three Gaussian components, one on the blue-shifted side and two in the red-shifted side. Table~\ref{tab:3gauss} summarizes the fitting results, and each component and the fitting results are shown in Figure~\ref{fig:gaussfit}. 
Except for a peak on the shoulder in ALMA2, the fitting with three Gaussians works well. 
We subtract component 1 from each line profile and apply the Hill5 model fitting as in Section~\ref{sec:hill5}. 

\begin{deluxetable*}{lcccccc}
\label{tab:3gauss}
\tabletypesize{\footnotesize}
\tablecaption{Summary of the three gaussian fitting}
\tablewidth{0pt}
\tablehead{
\colhead{}  & \colhead{Amplitude} & \colhead{Velocity center} & \colhead{sigma} \\ 
\colhead{} & \colhead{K} & \colhead{km\,s$^{-1}$} & \colhead{km\,s$^{-1}$}}
\startdata
ALMA2 comp.1 & 1.60$\pm$0.06  & -52.13$\pm$0.11 & 1.15$\pm$0.06 \\
ALMA2 comp.2 & 1.45$\pm$0.17  & -53.40$\pm$0.02 & 0.48$\pm$0.04 \\
ALMA2 comp.3 & 3.09$\pm$0.05  & -55.43$\pm$0.01 & 0.30$\pm$0.01 \\
ALMA18 comp.1 & 0.83$\pm$0.09  & -53.12$\pm$0.17 & 0.87$\pm$0.07 \\
ALMA18 comp.2 & 1.21$\pm$0.18  & -53.92$\pm$0.01 & 0.41$\pm$0.04 \\
ALMA18 comp.3 & 2.03$\pm$0.04  & -55.62$\pm$0.01 & 0.26$\pm$0.01 \\
\enddata
\end{deluxetable*}

\end{document}